\documentclass[trackchanges]{aastex701}

\usepackage{graphicx}
\usepackage{amsmath}
\usepackage{array}
\usepackage{enumitem} 
\usepackage[normalem]{ulem}
\usepackage{siunitx}
\usepackage{xcolor}
\usepackage{soul} 
\usepackage[normalem]{ulem}  
\usepackage{natbib}

\defcitealias{goulding09}{GA09}
\defcitealias{annuar25}{A25}
\defcitealias{dasyra08}{D08}

\begin{document}

\title{A Census of Active Galactic Nuclei Identified by the Mid-infrared [Ne\,{\sc v}] Line at $\boldsymbol{z \leq 0.025}$}

\author[gname=Sarah, sname='Yassir']{Sarah Yassir}
\affiliation{Department of Applied Physics, Faculty of Science and Technology, Universiti Kebangsaan Malaysia, 43600 UKM Bangi, Selangor, Malaysia}
\email{sarahyassir@gmail.com.my}

\author[orcid=0000-0003-0387-1429,gname=Adlyka, sname='Annuar']{Adlyka Annuar$^*$} 
\affiliation{Department of Applied Physics, Faculty of Science and Technology, Universiti Kebangsaan Malaysia, 43600 UKM Bangi, Selangor, Malaysia}
\email[show]{adlyka@ukm.edu.my}

\begin{abstract}
We present a census of local active galactic nuclei (AGN) at a redshift of \(z\leq 0.025\) selected using the high-ionization [Ne\,{\sc v}]\(\lambda14.32\,\mu\mathrm{m}\) emission line from the Infrared Database of Extragalactic Observables from Spitzer (IDEOS). We identify 103 sources with detected [Ne\,{\sc v}] emission, which we regard as AGN within the volume. This sample represents \(\sim 18\%\) of the galaxy population within this redshift range, consistent with AGN fractions derived using other selection techniques. We investigate the biases and properties of this [Ne\,{\sc v}]-selected AGN sample by comparing it with traditional AGN selection methods based on hard X-ray, optical, and mid-infrared colors. We find that our selection significantly misses AGN with underdeveloped narrow line regions {NLRs}, which account for approximately half of the AGN identified by NLR-independent methods. However, approximately $\sim10\%$ of our sample are undetected in optical diagnostics, while $\sim40\%$ are missed by hard X-rays and $\sim70\%$ by infrared continuum. Notably, $\sim 15\%$ of our AGN are missed by all classical methods, constituting a population of previously unidentified AGN revealed solely by the [Ne\,{\sc v}] emission line. Based on our analysis, we show that this line can efficiently select heavily Compton-thick and host-dominated AGN systems. Our analysis also yields mean bolometric luminosities of $\log(L_{{\rm bol}}/\mathrm{erg\,s^{-1}}) = 44.5 \pm 0.7$, black hole masses of $\log(M_{{\rm BH}}/M_\odot) = 7.3 \pm 0.6$, and Eddington ratio of $\lambda_{\mathrm{Edd}} = 0.15 \pm 0.11$. Our sample harbors AGN with comparable luminosities but systematically lower-mass black holes accreting at higher Eddington ratios than those in the hard X-ray–selected sample. This suggests that our AGN may represent local analogs of the rapidly growing SMBH population prevalent at cosmic noon.
\end{abstract}



\section{Introduction} 

Active Galactic Nuclei (AGN) represent one of the most energetic phenomena in the universe, powered by accretion onto supermassive black holes (SMBHs) at galactic centers. AGN are fundamental drivers of galaxy evolution \citep{aird12} and key contributors to the cosmic X-ray background and the accretion power budget of the universe \citep{comastri04, ueda14, ananna19, cox25}. A complete census of AGN is therefore essential for understanding black hole growth and its impact on host galaxies. However, obtaining this census is challenging because different observational methods select different incomplete AGN populations. Historically, AGN have been identified across the electromagnetic spectrum, with each waveband tracing distinct physical processes and suffering from unique selection effects. For example, although optical surveys are extensive and have generated large AGN catalogs \citep{veron10, zaw19}, they tend to miss AGN that are obscured by dust and gas in the torus and/or host galaxies that could suppress emission from the narrow-line region (NLR; \citealt{antonucci93, fiore00, goulding09, prieto21}). These obscured AGN have been shown to constitute a significant fraction of the local population ($\sim30 - 50\%$; \citealt{goulding11, ananna19, kim23, boorman25}). However, X-ray selection, which probes the AGN corona, can be contaminated by non-AGN sources such as ultra-luminous X-ray sources (ULX) and X-ray binaries (XRB), particularly at soft X-rays ($\leq 10$ keV, \citealt{shakura73, maccacaro88, kaaret17}). Hard X-rays ($\geq 10$ keV) are largely free from such contamination, but can still miss \textit{heavily} Compton-thick AGN with absorption column densities of $N_{\rm H} \geq 10^{25} \,\rm cm^{-2}$ \citep{annuar17, hatcher21}. Infrared diagnostics are less affected by dust obscuration, but can be severely contaminated by host-galaxy star formation \citep{hainline16fior, assef18}. Consequently, a complete AGN census remains fragmented when relying on these individual standard methods. Clearly, adopting a multiwavelength selection approach would result in a more complete sample, independent of AGN diagnostics at any wavelength. However, using a single waveband selection provides a simpler and more well-understood selection effect.

An alternative route to select AGN lies in mid-infrared (MIR) high-ionization “coronal” lines, such as [Ne\,{\sc v}]$\lambda14.32\,\mu\mathrm{m}$ produced in the NLR \citep{mignoli13, richardson14, amorim21, kynoch22}. Producing [Ne\,{\sc v}] requires photons with energies $\sim97.1$ eV, a condition rarely met outside AGN environments, making it a near-unambiguous AGN tracer \citep{weedman05, goulding09, tommasin10}. Even when [Ne\,{\sc v}] is produced by non-AGN activity, its luminosity is typically significantly weaker than that of AGN-powered emission \citep{smith09-sn, pottasch09, tarantino24}. Furthermore, MIR emission penetrates dust more readily than optical or soft X-ray light, making it sensitive to heavily obscured systems. Several studies using the \textit{Spitzer} Space Telescope \citep{werner04-spitzer} have empirically confirmed the diagnostic power of this line. For example, \citet{goulding09} (hereafter \citetalias{goulding09}) formed a sample of AGN within 15 Mpc using the [Ne\,{\sc v}] line, and showed that about half of their sample are not detected in optical spectroscopic diagnostics. This highlights the [Ne\,{\sc v}] line abilities to recover AGN that are missed in optical surveys mainly due to host galaxies obscuration. In addition, a significant amount of the AGN discovered by \citetalias{goulding09} without prior optical AGN signatures are found to be heavily obscured, and even Compton-thick with column densities exceeding $10^{24} \,\rm cm^{-2}$ \citep{annuar17, annuar20, annuar25}. These studies show the importance of [Ne\,{\sc v}] as a heavily obscured AGN tracer. However, in heavily obscured AGN, dense circumnuclear material can suppress or prevent the formation of an extended NLR \citep{imanishi06}, leading to weak or absent [Ne\,{\sc v}] emission. Consequently, [Ne\,{\sc v}] selection can miss such deeply embedded AGN with under-developed NLR \citep{maiolino03, imanishi06, imanishi08}, a limitation shared by other NLR-dependent diagnostics such as optical spectroscopic selection.

Despite this, [Ne\,{\sc v}] can still provide an independent AGN selection method that can complement traditional methods. Beyond mere detection, [Ne\,{\sc v}] offers a quantitative pathway to estimate fundamental AGN properties that are critical to understanding their evolution and impact. This highly ionized line has evidence that it actually detects the accretion-powered emission of AGN \citep{weaver10, prieto22, bierschank24}. Accurately measuring black hole masses (\(M_{\rm{BH}}\)) and bolometric luminosities (\(L_{\text{bol}}\)) forms the basis for studying AGN demographics and their coevolution with host galaxies \citep{magorrian98}. With both $M_{\rm{BH}}$ and \(L_{\text{bol}}\) estimated, the accretion rate, expressed as the Eddington ratio (\(\lambda_{\text{Edd}} = L_{\text{bol}}/L_{\text{Edd}}\)), can be derived. This parameter is crucial for probing AGN physics, as it distinguishes between radiatively efficient (\(\lambda_{\text{Edd}} \gtrsim 0.01\)) and inefficient accretion modes and governs feedback processes that regulate gas supply and obscuration \citep{ricci17-nature, gupta25}. 

The use of NLR emission lines as black hole mass proxies is motivated by several observed correlations between AGN properties and $M_{\text{BH}}$. For example, studies have shown that NLR line widths are comparable to stellar velocity dispersions \citep{nelson96, greene05}, suggesting that NLR kinematics may trace the gravitational potential of the host galaxy bulge and thus relate to the well-known $M_{\text{BH}}- \sigma^*$ relation \citep{king03, faucher12, tart26}. In addition, correlations have also been reported between the luminosities of certain NLR emission lines and $M_{\text{BH}}$, such as the relation involving [O\,{\sc iii}]{$\lambda$}5007{$\si{\angstrom}$} luminosity and $M_{\text{BH}}$ \citep{harms94, nelson00, shields04, wang09}. These luminosity$-M_{\text{BH}}$ relations are primarily empirical, but they provide a useful observational scaling for estimating black hole masses when direct dynamical measurements are unavailable. In fact, \citet{dasyra08} (hereafter \citetalias{dasyra08}) have shown relations between [Ne\,{\sc v}]$\lambda14.32\,\mu\mathrm{m}$ luminosity against \(L_{\text{bol}}\) and $M_{\rm{BH}}$ obtained using the reverberation-mapping technique. They Using a sample of AGN with reverberation-mapped $M_{\rm{BH}}$ measurements, \citetalias{dasyra08} found that [Ne\,{\sc v}] luminosity correlates well with both $M_{\rm{BH}}$ and \(L_{\text{bol}}\) with scatters $\sim$0.46 dex, a precision comparable to the renowned \(M_{\text{BH}}-\sigma^*\) relations \citep{ferrarese00, kormendy13}. Subsequent work by \citet{spignolio22} confirmed a similarly tight correlation between [Ne\,{\sc v}] luminosity and \(L_{\text{bol}}\) (scatter $\sim$0.3 dex). These relatively low-scatter relations confirm that [Ne\,{\sc v}] is a reliable proxy for the intrinsic AGN power.

In this work, we compile and analyze an AGN sample identified solely by the presence of [Ne\,{\sc v}]$\lambda14.32\,\mu\mathrm{m}$ using the \textit{Spitzer Space Telescope}. Our primary goal is to understand the biases and fundamental properties of this AGN sample, mainly \(L_{\text{bol}}\), $M_{\rm{BH}}$, and \(\lambda_{\text{Edd}}\), thus mapping the demographics of AGN selected by this high-ionization tracer. This paper is organized as follows. Section~\ref{sec:sample} provides a description of how we constructed the sample and examines its completeness and reliability. We then investigate the biases and properties of our [Ne\,{\sc v}] sample by examining how these sources are identified using traditional selection methods at multiple wavelengths in Section~\ref{sec:multiwavelength}. In Section~\ref{sec:agn-properties} we discuss the physical properties of the AGN, including bolometric luminosities, black hole masses, and accretion rates. Finally, Section~\ref{conclusions} provides a discussion and summary of our principal findings and their implications. Throughout this work, we adopt a concordance cosmology with $\Omega_M = 0.286$, $\Omega_{\Lambda} = 0.714$, and $H_0 = 69.6$ km s$^{-1}$ Mpc$^{-1}$. All fit uncertainties are quoted at 90\% confidence levels unless otherwise explicitly stated.

\section{[Ne\,{\sc v}] AGN Sample Selection} 
\label{sec:sample}

In this work, we used the Infrared Database of Extragalactic Observables from \textit{Spitzer} (IDEOS, \citealt{spoon22}) catalog to construct our  AGN sample. We briefly describe the galaxy selection here and refer the reader to \citet{spoon22} and the previous paper on IDEOS by \citet{hernan16} for more comprehensive details. The IDEOS catalog comprises 3335 extragalactic galaxies observed with the Spitzer Infrared Spectrograph (IRS), derived from the Cornell Atlas of Spitzer/IRS Sources (CASSIS; \citealt{Lebouteiller11}) repository. This catalog provides homogeneous measurements of spectral features and continuum flux densities obtained through fitting and decomposition of IRS low-resolution spectra ($R =$ 60–120). For this study, we used high ionization [Ne\,{\sc v}] as an indicator of the presence of AGN in galaxies within IDEOS. Based on this, we found 371/3335 ($11.12 \pm 5.70\%$) AGN in the catalog that have [Ne\,{\sc v}] line detection. Among these, 18 sources have multiple [Ne\,{\sc v}] measurements. Therefore, for these objects, we calculated the mean flux values to obtain a single measurement per source.

\begin{figure*}
    \centering
	\includegraphics[width=0.5\textwidth]{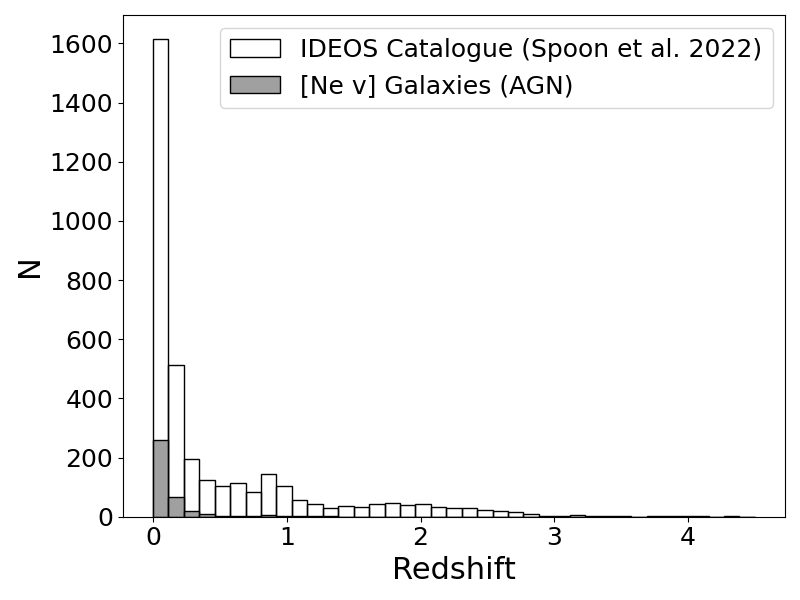}\\
    \includegraphics[width=1.0\textwidth]{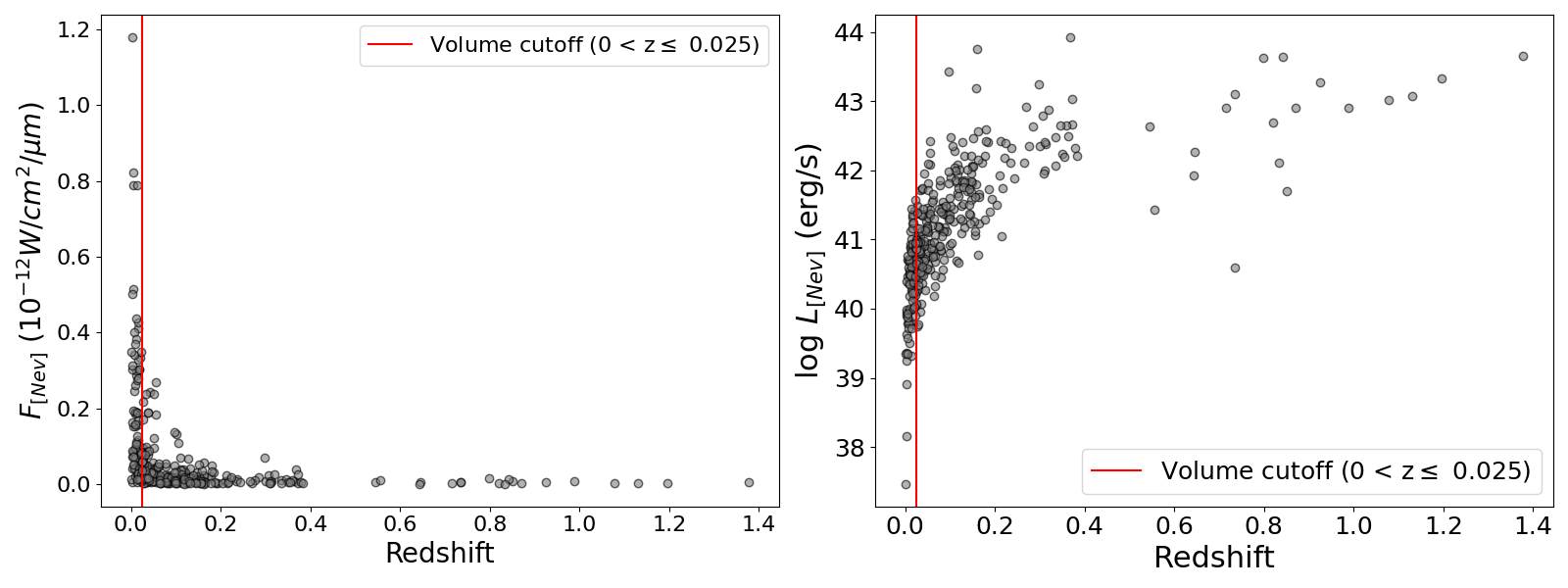} \\
    \caption{\textit{Top}: Redshift distribution of the full IDEOS sample (black solid line), and [Ne\,{\sc v}]-selected galaxies (gray shaded). The [Ne\,{\sc v}] contributes to $\sim 11\%$ of the whole IDEOS catalog \citep{spoon22}. The [Ne\,{\sc v}] AGN sample shows a strong concentration at low redshifts while diminishing exponentially indicating some selection biases of \textit{Spitzer}/IRS instrument. \textit{Bottom:} Distribution of [Ne\,{\sc v}]$\lambda$14.32 \micron\ flux and luminosity against redshift. We empirically determined a flux completeness limit by identifying the maximum redshift boundary. The vertical line marks the boundary for the AGN complete sample (z = 0.025).}
    \label{fig:sample}
\end{figure*}

Figure~\ref{fig:sample} (top) shows the number of sources in the IDEOS catalog as a function of redshift. In this figure we can see that many galaxies with [Ne\,{\sc v}] line detection dominate the low redshift region, most likely due to \textit{Spitzer}/IRS’ sensitivity limits (0.4 mJy for low-resolution modules in 512 seconds of integration). The steep decline at higher redshift in population density indicates flux and luminosity bias of [Ne\,{\sc v}] detection at a certain redshift value. Figure~\ref{fig:sample} (bottom) shows the [Ne\,{\sc v}] flux and luminosity distributions across redshift. The plots show a significant drop in sources at $z \sim 0.05$, indicative of observational selection effects where fainter objects at higher redshifts are missed due to the flux limitation of \textit{Spitzer}. This might lead to systematic biases and an inaccurate representation of the AGN population that would lead to distorted conclusions. Therefore, it is crucial to construct a volume-limited sample to be able to make measurements with a reliable data set.  

\subsection{Sample Completeness and Reliability}
\label{sample-completeness}

\subsubsection{IDEOS Sensitivity limit}
\label{sample-2.1.1}

To construct a volume-limited sample, we employed a systematic percentile-based approach that determines both luminosity and redshift completeness limits. We identified a minimum luminosity threshold of $L_{\min} = 2.69 \times 10^{39}$ erg s$^{-1}$ (98th percentile of the observed [Ne\,{\sc v}] luminosity distribution) and a corresponding flux limit of $F_{\min} = 1.91 \times 10^{-15}$ erg s$^{-1}$ cm$^{-2}$ (98th percentile of the measured [Ne\,{\sc v}] flux distribution). The maximum redshift for \textit{Spitzer}'s nominal sensitivity limit was derived by solving the luminosity-distance relation. This calculation determines the farthest distance at which a source with luminosity $L_{\min}$ would still be detectable given the flux limit $F_{\min}$. Although this method identifies a minimum luminosity threshold and a corresponding flux limit, we intentionally applied only the redshift cutoff to define our final volume-limited sample. This approach was chosen for a critical scientific reason where applying both the $L_{\min}$ threshold and the redshift cutoff would exclude low-redshift AGN with luminosities below $L_{\min}$, artificially restricting our sample's dynamic range and potentially biasing our analysis toward only the most luminous systems. Since our primary goal was to study the AGN population within a well-defined volume rather than at a specific luminosity threshold, we preserved all sources within the redshift boundary, regardless of their individual luminosities. This ensured that our volume-limited sample captures the full diversity of AGN properties present in the local universe while maintaining complete detectability within the defined cosmological volume. We found a completeness redshift of z = 0.025. This represents the redshift limit beyond which our sample would become incomplete due to sensitivity constraints of the \textit{Spitzer} observations.

The final volume-limited sample comprises 103 of 575 galaxies within \(z \leq 0.025\) (\(D \lesssim 105 \ \mathrm{Mpc}\)), providing a complete sample of AGN within this redshift-limited volume, free from incompleteness biases that affect flux-limited samples. The complete sample is listed in Table~\ref{appA} (see Appendix) and represents \(17.91 \pm 6.95\%\) of the sources of the IDEOS parent catalog within this volume. This AGN fraction is consistent, within the uncertainty limits, with previous [Ne\,{\sc v}] study by \citetalias{goulding09} who obtained a fraction of $27^{+8}_{-6}\%$. In addition, our population is very close to the optical yields of an all-sky optical spectroscopy survey by \citealt{zaw19}, which found an AGN fraction of $19.04 \pm 7.11\%$. Our fraction is also comparable to radio AGN yields ($15.12 \pm 6.49\%$, \citealt{delvecchio17}), although radio traces jet-dominated systems \citep{best12} while [Ne\,{\sc v}] probes accretion-powered emission. However, we found significantly higher AGN rates than infrared continuum selections ($0.18\%-2.80\%$, \citealt{secrest15, assef18}), highlighting the advantage of [Ne\,{\sc v}]’s high ionization potential over broadband mid-infrared colors, which suffer from strong host galaxy contamination \citep{assef13}. Our fraction is also notably lower than X-ray selections ($34\%-44\%$, \citealt{she17, oh18-swiftbat}), underscoring the complementary nature of the two methods in which X-rays efficiently detect unobscured and moderately obscured AGN but miss heavily Compton-thick systems {($N_{\rm H} \geq 10^{25} \,\rm cm^{-2}$), while [Ne\,{\sc v}], less affected by obscuration, can recover such missing populations, but is biased against those with under-developed NLR.

The [Ne\,{\sc v}] luminosities of our sample span a wide dynamic range of 4 orders of magnitude, from \(10^{37} \ \mathrm{erg \ s^{-1}}\) to \(10^{41} \ \mathrm{erg \ s^{-1}}\), with a mean of \(5.2 \times 10^{40} \ \mathrm{erg \ s^{-1}}\). We find no strong correlation between [Ne\,{\sc v}] luminosity and redshift within our narrow redshift window (\(z \leq 0.025\)), confirming that the observed luminosity range primarily reflects intrinsic AGN power rather than distance effects. This luminosity range is comparable to those reported in previous studies of local AGN samples (e.g. \citetalias{goulding09, dasyra08}), validating our sample as representative of the local AGN population selected by [Ne\,{\sc v}].

\subsubsection{Contribution from Non-AGN Sources}
\label{sample-2.1.2}

In this section, we investigate whether the [Ne\,{\sc v}] line detected could be contributed by non-AGN sources such as stellar processes and ULXs. Figure~\ref{fig:[nev]-line-ratio} presents the diagnostic diagram of $\log$([Ne\,{\sc v}]/[Ne\,{\sc ii}]) versus $\log$([Ne\,{\sc iii}]/[Ne\,{\sc ii}]). This diagram can be used to distinguish between [Ne\,{\sc v}] emission dominated by AGN and star formation activities \citep{goulding09, lamassa12, cleri23}. We applied the empirical boundaries of $\log$([Ne\,{\sc v}]/[Ne\,{\sc ii}]) $\geq$ -1 by \citet{inami13} and $\log$([Ne\,{\sc iii}]/[Ne\,{\sc ii}]) $\geq$ 0 by \citet{melendez08} to define an AGN-dominated region. Based on the figure, more than half of our sample is located within the AGN-dominated region. However, there are a significant number of sources outside of this region. The majority of these lie in the star-forming dominated region defined by \citet{inami13}. Interestingly, seven out of eight sources in our sample that have luminosity below the minimum threshold derived in Section~\ref{sample-2.1.1} are located within the AGN-dominated region. This indicates that the [Ne\,{\sc v}] line detected is most likely the result of AGN activity. The remaining source (i.e., NGC 6328) lies outside of this region but very close to the AGN-dominated boundary.

In addition, we also plot empirical line ratios for available Wolf-Rayet (WR) galaxies (\citet{tarantino24}; \citetalias{goulding09}). These sources exhibit high $\log$([Ne\,{\sc iii}]/[Ne\,{\sc ii}]) but consistently low $\log$([Ne\,{\sc v}]/[Ne\,{\sc ii}]) (\( \lesssim -1\)). This is consistent with photoionization models of WR populations, which show that even the hottest stellar spectra cannot produce $\log$([Ne\,{\sc v}]/[Ne\,{\sc ii}]) $>$ -1 \citep{schaerer99}. The WR galaxies therefore occupy a distinct region of the diagram and do not intrude into the parameter space of our sample. None of our samples are located near the WR region, indicating that they are unlikely to have a significant contribution from this source.

Next, we compared our [Ne\,{\sc v}] luminosity threshold, i.e. \(L_{\min} = 2.69 \times 10^{39}\) erg s\(^{-1}\) (Section~\ref{sample-2.1.1}) to observed and modeled [Ne\,{\sc v}] luminosities from known non-AGN sources. We found that all such sources fall well below our limit. For example, the ULX observed in SBS 0335‑052E has \(\log L_{\rm [Ne\,v]} \sim 36\) erg s\(^{-1}\) \citep{mingozzi25}, while ULX X-1 in NGC 6946 was found with undetected [Ne\,{\sc v}] luminosity but with a derived upper limit of \(\log L_{\rm [Ne\,v]} \lesssim 38.5\) erg s\(^{-1}\) \citep{berghea12}. In addition, supernova remnants and planetary nebulae have been detected with \(L_{\rm [Ne\,v]}\) of \(\log \sim 33\) erg s\(^{-1}\) \citep{smith09-sn, pottasch09}. Furthermore, in extreme stellar environments such as Wolf‑Rayet nebulae and H\,{\sc ii} galaxies, they are seen to exhibit \(\log L_{\rm [Ne\,v]} \sim 36-37\) erg s\(^{-1}\) and \(< 38\) erg s\(^{-1}\), respectively \citep{tarantino24, pereira10}. Fast radiative shock models also predict \(\log L_{\rm [Ne\,v]} \lesssim 38.5\) erg s\(^{-1}\) \citep{dopita95, izotov21}.

\begin{figure*}
    \centering
	\includegraphics[width=0.7\textwidth]{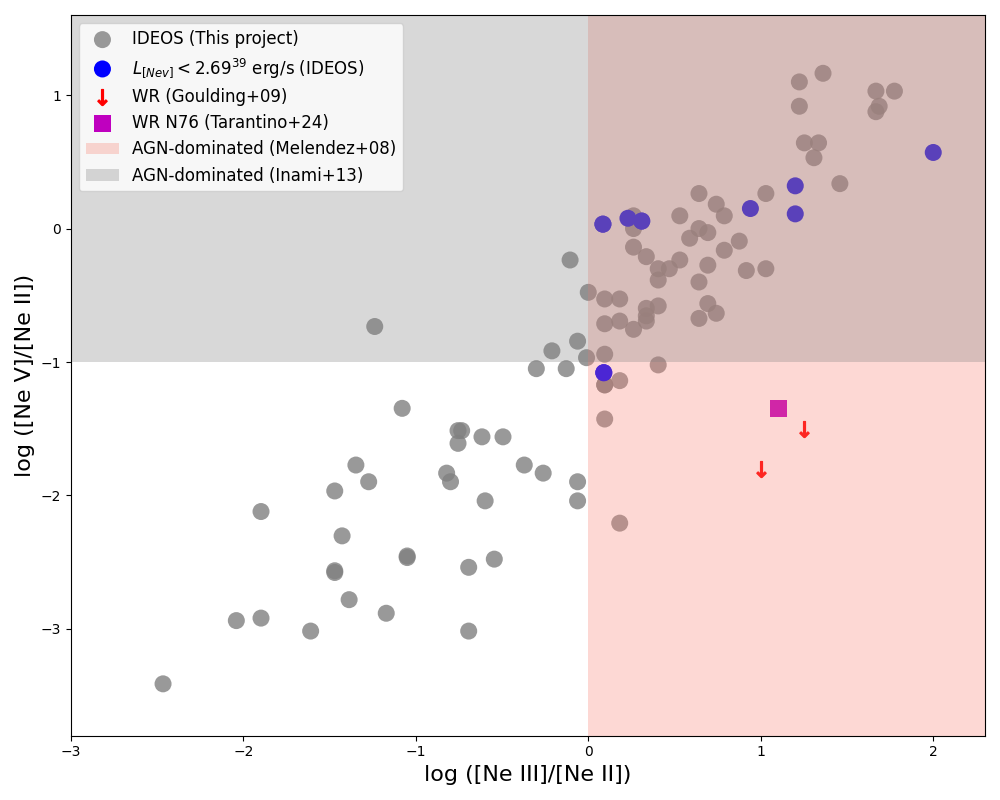}\\
    \caption{Diagnostic diagram of $\log$([Ne\,{\sc v}]/[Ne\,{\sc ii}]) vs. $\log$([Ne\,{\sc iii}]/[Ne\,{\sc ii}]) for our [Ne\,{\sc v}]-selected AGN sample, using empirical AGN boundaries from \citet{inami13} (gray shaded region, $\log$([Ne\,{\sc v}]/[Ne\,{\sc ii}]) $\geq$ -1) and \citet{melendez08} (red shaded region, $\log$([Ne\,{\sc iii}]/[Ne\,{\sc ii}]) $\geq$ 0). Blue circles mark sources below our luminosity threshold (Section~\ref{sample-2.1.1}); seven of these lie inside the AGN region, confirming their AGN origin, while NGC 6328 lies just outside but close to the AGN boundary. Although many sources fall outside the AGN region, their high [Ne\,{\sc v}] luminosities (above those of any known non-AGN source) confirm that all objects in the sample are genuine AGN, with star formation contributing to the line ratios of some objects. We also include Wolf–Rayet (WR) sources for comparison. Line ratios from \citet{tarantino24} (magenta squares) were measured in a WR nebula, while two WR galaxies from \citetalias{goulding09} with [Ne\,{\sc v}] upper limits are shown as red arrows.}
    \label{fig:[nev]-line-ratio}
\end{figure*}

Collectively, non-AGN sources were found/expected to have [Ne\,{\sc v}] luminosities at least an order of magnitude lower than our threshold. This further supports the notion that the majority of the [Ne\,{\sc v}] emission detected in our sample originates from AGN activity. We therefore conclude that [Ne\,{\sc v}] selection provides a reliable tracer of AGN in the local universe.

\subsubsection{Under-developed Narrow Line Region}
\label{sample-2.1.3}

Although [Ne\,{\sc v}] is a good proxy for AGN, its origin in NLR means that we could miss the deeply embedded AGN with under-developed NLR \citep{imanishi08}. In this section, we investigate whether our [Ne\,{\sc v}] sample is affected by this selection effect by comparing all IDEOS galaxies within our redshift limit (i.e. 0.025) to the AGN sample identified via independent methods that are not affected by this selection bias.

For example, the mid-IR continuum selection method, such as the WISE color-color selection used in the AllWISE AGN catalog \citep{secrest15}, is not affected by the NLR because it traces thermal dust emission from the circumnuclear material \citep{netzer93, honig12}. The AllWISE AGN catalog is an all-sky sample containing $\sim$1.4 million AGN that were identified as AGN by two-color infrared photometric selection criteria. The catalog was tested for complete detection and resulted in $\geq 80\%$ completeness for AGN with magnitude, $R \leq 19$. We therefore cross-matched the IDEOS galaxies (within $z \leq 0.025$) with the AllWISE AGN catalog to identify AGN selected by their mid-IR color selection. This yields 46 AGN. Of these, 22 sources (47.8 ± 9.7\%) are not identified as AGN based on [Ne\,{\sc v}] detection. This indicates that their [Ne\,{\sc v}] are weak or absent, suggesting that their NLRs might be under-developed.

In addition, hard X-ray ($\geq 10$ keV) selection probes the accretion process directly via inverse Compton scattering in the corona, located just outside the black hole. This high-energy emission is highly penetrating, capable of emerging through obscuration (up to $N_{\rm H} \sim 10^{25} \,\rm cm^{-2}$) that would suppress the NLR, thus revealing AGN activity even in deeply embedded nuclei \citep{ibar07, hickox18}. The \textit{Swift}/BAT 105 month catalog \citep{oh18-swiftbat} is a uniform hard X-ray all-sky survey performed in the first 105 months of observations using the Burst Alert Telescope (BAT) imager within the \textit{Swift} observatory. BAT is a highly sensitive instrument that detects 90\% of the sky in the $14-195$ keV band. We cross-matched the galaxies in the IDEOS sample within the redshift limit (i.e., 0.025), with the \textit{Swift}/BAT 105-month catalog \citep{oh18-swiftbat}. We found 104 IDEOS galaxies with X-ray counterparts. Of these, 44 (42.3 ± 9.5\%) do not have [Ne\,{\sc v}] emission line, suggesting that their NLR are under-developed. This value is consistent with the mid-IR continuum findings.

These analyzes indicate that our [Ne\,{\sc v}] selection misses a significant fraction of AGN, approximately 45\% which likely represents AGN with under-developed NLRs. Therefore, this AGN selection is highly biased against these sources. Nevertheless, it remains a robust tracer for AGN accretion activity when detected (Section~\ref{sample-2.1.2}).

\section{Multiwavelength Detection}
\label{sec:multiwavelength}

In this section, we cross-match our sample with traditional AGN selection techniques in the X-ray, optical, and mid-infrared continuum to understand the biases and properties of our [Ne\,{\sc v}] sample. Table~\ref{detection_rates} presents the summary of our result and a more detailed result is presented in Table~\ref{appA} (see the Appendix).

\subsection{X-ray}
\label{subsec:xray-detection}

\begin{figure}
    \centering
    \includegraphics[width=0.7\columnwidth]{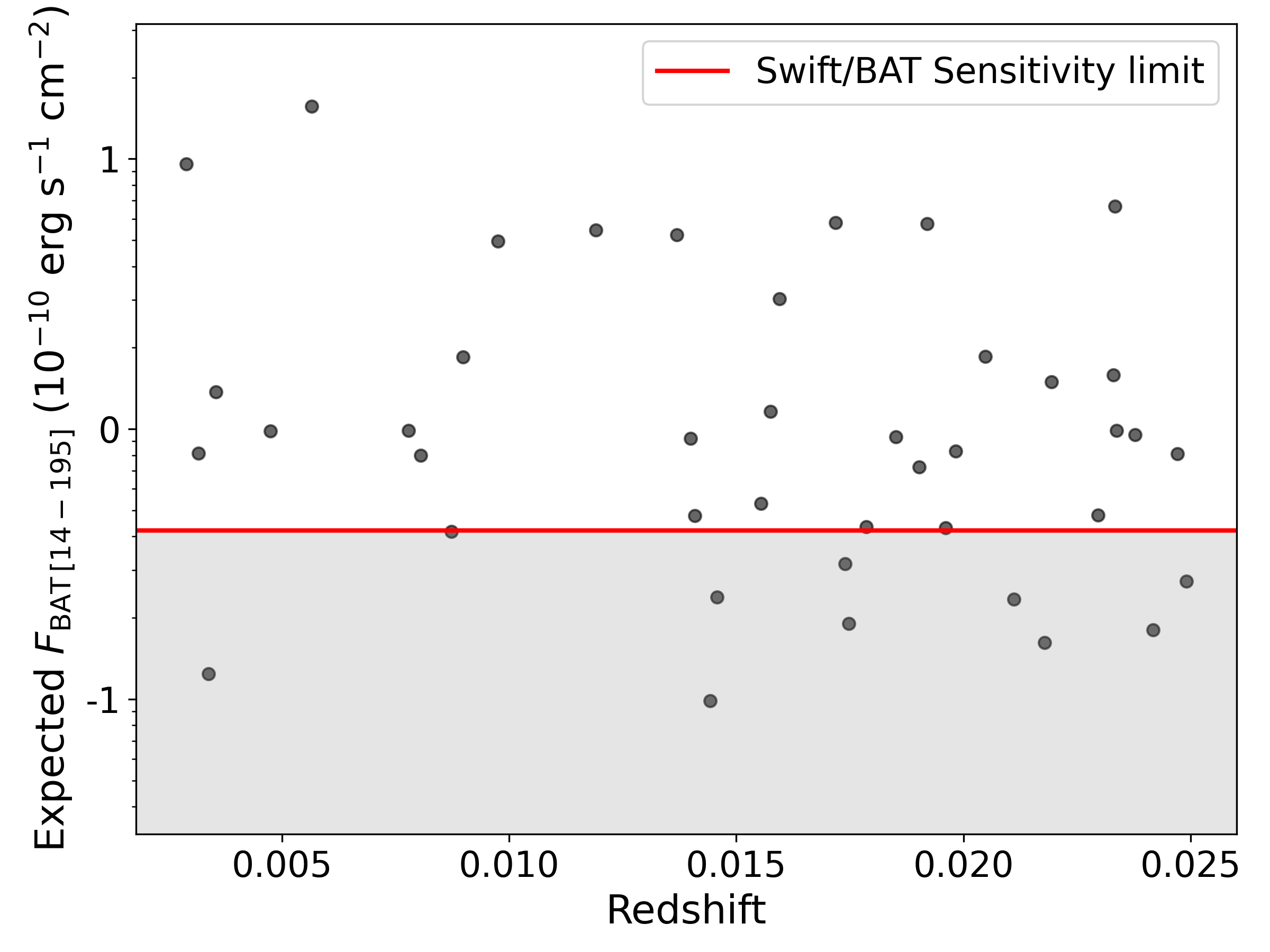}
    \caption{Estimated \textit{Swift}/BAT fluxes for the X-ray–undetected ($\sim$40\%) AGN inferred from the [Ne\,{\sc v}]–BAT relation as a function of redshift. The red line and shaded region denote the $5\sigma$ \textit{Swift}/BAT sensitivity limit. We found that out of 41 AGN undetected by \textit{Swift}/BAT, 10 have expected BAT flux below the BAT sensitivity limit flux while 31/41 sources are above the limit indicating that they could have been detected by \textit{Swift}/BAT. This indicates that they may be heavily Compton-thick AGN.}
    \label{fig:weaver}
\end{figure}

Hard X-ray selection ($E \geq$ 10 keV) provides a robust method for identifying AGN. At these energies, emission is less affected by obscuration (up to $N_{\rm H} = 10^{25} \ \text{cm}^{-2}$) and is largely free from contamination by normal stellar processes. Although XRBs and ULXs can emit hard X-rays, their X-ray spectrum typically turn down around $E \sim 10$ keV \citep{bachetti13, annuar15, brightman18} and their luminosities typically reach $\lesssim 40$ erg s$^{-1}$ \citep{kim04-lmxb, mineo12-hmxb, ueda14, aird15}.

We therefore evaluated our [Ne\,{\sc v}]-selected AGN using data from the \textit{Swift}/BAT all-sky survey. Given the high energy operating range of \textit{Swift}/BAT, it is largely insensitive to emission from non-AGN sources, making it an ideal tool for cross-matching our AGN sample. We cross-matched our sample with the \textit{Swift}/BAT 105-Month catalog \citep{oh18-swiftbat}. Using a $5\arcmin$ matching radius (consistent with \textit{Swift}/BAT's PSF FWHM) and requiring S/N $> 4.8$ for ${\approx}99\%$ reliability \citep{topcat}, we identified X-ray counterparts for our AGN. Based on this, we found that 62 of 103 of our [Ne\,{\sc v}] sources ($60.2 \pm 8.9\%$) are detected in BAT, which means that nearly 40\% of our AGN are not detected by hard X-ray selection. This result is consistent with \citetalias{annuar25}, which determined that only $47 \pm 9\%$ of their [Ne\,{\sc v}] AGN are detected in hard X-rays by \textit{Swift}/BAT.

Next, we investigated whether the \textit{Swift}/BAT non-detections are due to BAT sensitivity constraint or AGN obscuration. We derived the expected BAT fluxes for these sources using the [Ne\,{\sc v}]–BAT flux relation from \citet{weaver10}. We then compared these expected fluxes with the \textit{Swift}/BAT $5\sigma$ sensitivity limit. Figure~\ref{fig:weaver} shows the distribution of expected BAT fluxes versus redshift along with the BAT detection threshold. Based on this figure, we found that of the 41 AGN in our sample not detected by BAT, 10 have expected fluxes below the $5\sigma$ sensitivity limit and would not be reliably detected by the survey. The remaining 31 sources have expected fluxes above the detection threshold, indicating that they should be detectable by \textit{Swift}/BAT. This may suggest that their non-detection is likely due to heavy obscuration, specifically \textit{heavily} Compton-thick column densities ($N_{\mathrm{H}} \gtrsim 10^{25}$ cm$^{-2}$), which even the hard X-ray telescope would not be able to see. Therefore, we estimate that $30.10 \pm 8.31\%$ (31/103) of our [Ne\,{\sc v}] sample is heavily Compton-thick AGN. This fraction agrees with that predicted by population synthesis modeling by, for example, \citet{ueda14} and \citet{ananna19}, who assumed that heavily Compton-thick AGN ($\log N_{\mathrm{H}} = 25-26$ cm$^{-2}$) have the same number density as moderate Compton-thick AGN ($\log N_{\mathrm{H}} = 24-25$ cm$^{-2}$). This assumption leads to a heavily Compton-thick fraction of $25 \pm 4.5\%$ at $z<0.1$, which is consistent with our independently measured fraction of $30.10 \pm 8.31\%$. Extending this logic to \citet{ricci15}, their observed moderate Compton-thick fraction of $27 \pm 4\%$ would imply a similar heavily Compton-thick population. These findings further support our conclusion that the BAT-undetected [Ne\,{\sc v}] population consists predominantly of heavily Compton-thick AGN.

\subsection{Infrared}
\label{subsec:infrared}

Mid-infrared color can select AGN based on their characteristic hot dust emission from the tori, providing a complementary approach to identify obscured systems and star forming galaxies \citep{u22, derosa23}. We examined the performance of \textit{Wide-field Infrared Survey Explorer (WISE}) color-color criteria to recover our [Ne\,{\sc v}]-selected AGN. We applied mid-infrared color selection using \textit{WISE} photometry according to the \citet{mateos12} criteria of $W1 - W2 \geq 0.5$, $W2 - W3 \geq 2.5$ and an AGN locus ($y' = 0.315 \times x'$). We retrieved the data for our AGN, which required an SNR of $\geq 5$ in the W1, W2, and W3 bands and excluded sources with artifacts. We found that 88/103 of our sample meet the quality criteria and proceed to use them in investigating the efficiency of mid-infrared AGN color selection. 

\begin{figure}
\centering
	\includegraphics[width=0.7\columnwidth]{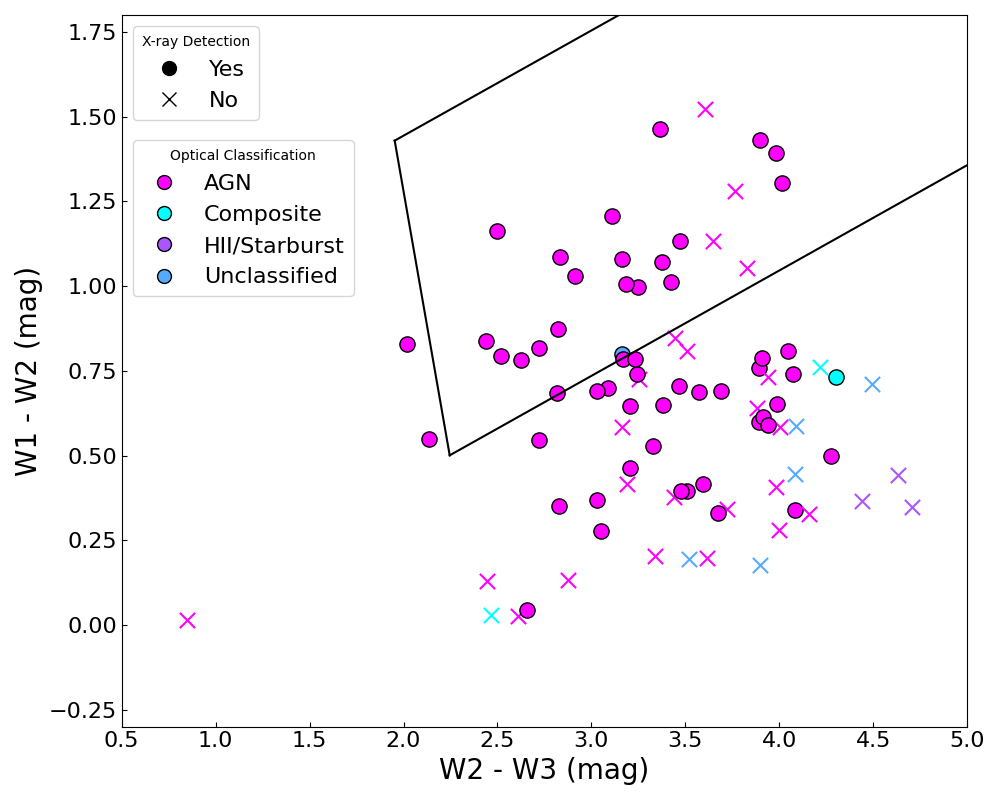}
    \caption{Mid-infrared color–color diagram based on \textit{WISE} photometry. The solid wedge delineates the AGN selection region defined by \citet{mateos12}. Marker shapes indicate X-ray classifications, while colors denote optical classifications. Approximately $70\%$ of the [Ne\,{\sc v}]-selected AGN lie outside the canonical \textit{WISE} selection wedge and exhibit systematically bluer W1$-$W2 colors than typical AGN populations, suggesting that their mid-infrared emission is dominated by stellar processes rather than by luminous, dust-obscured AGN activity.}
    \label{fig:wise}
\end{figure}

Figure~\ref{fig:wise} presents the WISE color-color diagram for our [Ne\,{\sc v}] sample. Based on this figure, only \(29.55 \pm 9.53\%\) (26/88) of our sources reside within the AGN wedge (\citealt{mateos12}). The majority of our sample falls outside of this wedge, displaying systematically bluer \(W1-W2\) colors. The distribution reveals distinct sub-populations. A subset of sources in the upper-left region (red \(W1-W2\), blue \(W2-W3\)) may indicate AGN with compact tori or minimal cool dust emission. In contrast, the significant population in the lower-right region (blue \(W1-W2\), red \(W2-W3\)) is consistent with strong contamination from host-galaxy stellar light and PAH emission \citep{stern12, asmus20}. This distribution demonstrates that while the [Ne\,{\sc v}] line robustly identifies AGN activity, the broadband MIR colors of most sources in our sample are dominated by host galaxy light, which dilutes the AGN's characteristic hot dust signature.

Our WISE color detection rate ($29.55 \pm 9.53\%$) exceeds the $14.35 \pm 6.35\%$ found by \citet{glikman18} in an optical parent sample. The higher rate stems from our use of a [Ne\,{\sc v}]-selected AGN sample, where all sources have confirmed nuclear activity, compared to their mixed galaxy sample, where many objects lack an AGN. This comparison highlights a well-documented limitation where broadband mid-IR color selection is strongly biased against AGN whose nuclear continuum is diluted by host-galaxy stellar emission. 

Studies combining optical and IR photometry consistently show that a significant population of spectroscopically confirmed AGN, especially at lower luminosities, reside outside of the standard selection wedges (e.g. \citealt{asmus20}). This bias is worsened by the all-sky photometric nature of the WISE survey, which has a lower sensitivity and resolution compared to pointed instruments such as \textit{Spitzer}/IRS ($R \sim 60-600$). The impact of this bias is starkly evident in our [Ne\,{\sc v}]-selected AGN sample. Here, $\sim 70\%$ of the AGN identified via this high-excitation nuclear line evades detection by standard WISE color selection. This result shows that while broadband mid-IR diagnostics are efficient for finding unobscured, luminous AGN with dominant hot dust emission, they are inefficient for identifying AGN in host-dominated systems where the nuclear mid-IR continuum is subdominant.

\subsection{Optical}
\label{subsec:optical}

\begin{figure}
\centering
\includegraphics[width=0.7\columnwidth]{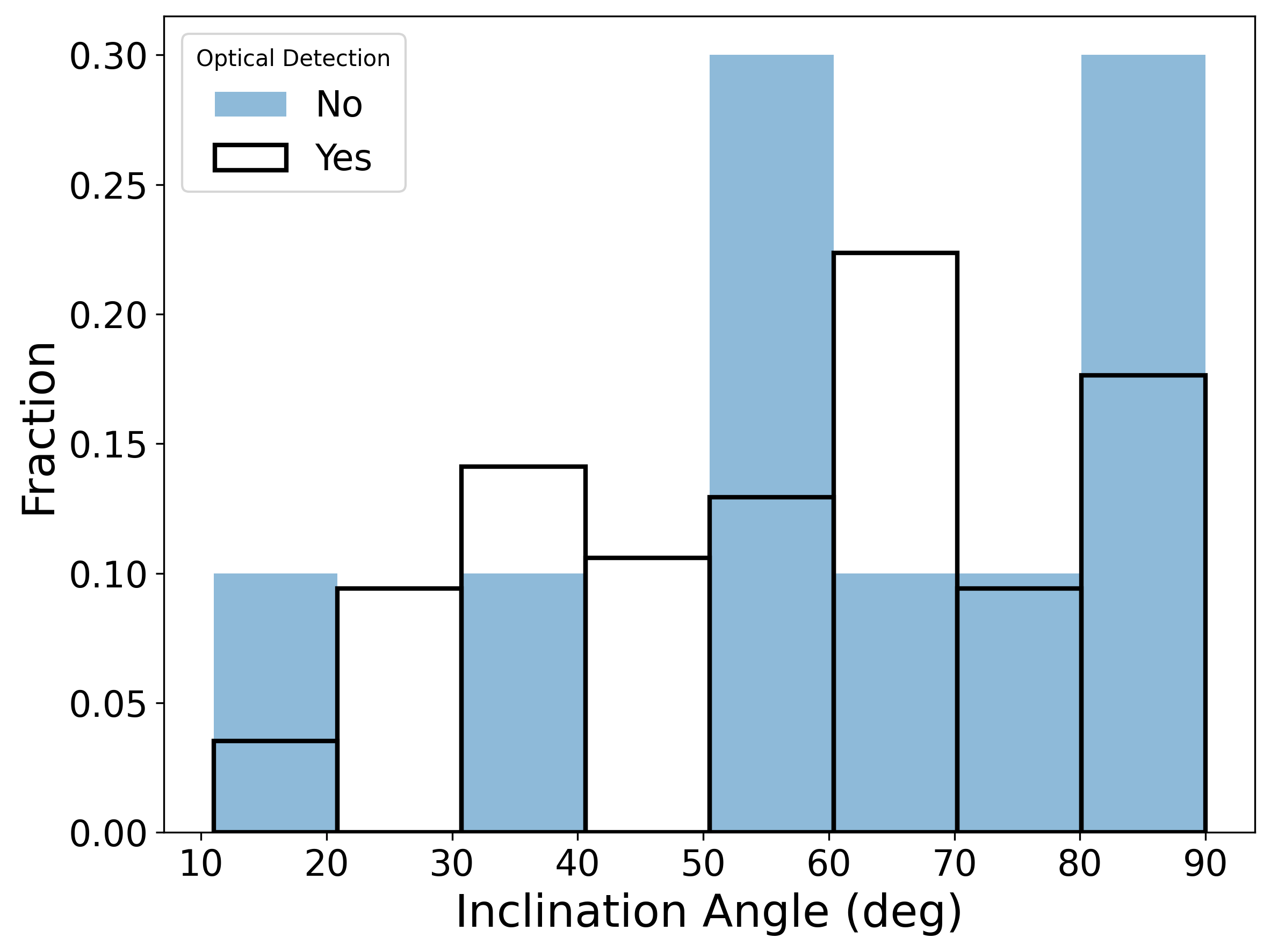}
\caption{Normalized histogram showing host galaxy inclination angle distributions for optically detected (red outline) and unidentified AGN (blue shaded). Mean inclinations are $57.8^\circ \pm 21.5$ for optical and $57.0^\circ \pm 21.4$ for non-optical AGN. A Kolmogorov–Smirnov test between the two distributions shows no significant difference ($D_{\rm KS} = 0.21$, $P_{\rm KS} = 0.75$), indicating that galaxy inclination does not affect optical detection rates.}
\label{fig:inclination}
\end{figure}

Optical diagnostics using emission line ratios provide the most established method for AGN identification, though they can be compromised by host galaxy obscuration. We assessed the properties of our [Ne\,{\sc v}] sample using optical AGN selection methods.

We identified our AGN in the optical using nuclear activity classifications from the NASA/IPAC Extragalactic Database (NED){\footnote{NED IPAC can be accessed through the following website: https://ned.ipac.caltech.edu/.}}. Based on this comprehensive search, we found that 7 out of 103 sources lack optical classifications. Of the remaining 96 sources, 86 are classified as AGN, corresponding to a detection rate of 86/96 ($89.58 \pm 6.11\%$). This selection yielded the highest detection rate among the three multi-wavelength methods we examined. The remaining 10/96 optically unidentified sources consist of 4 H\,{\sc ii}/starburst galaxies ($4.17 \pm 4.00\%$), 3 LINERs ($3.13 \pm 3.48\%$) and 3 composite sources ($3.13 \pm 3.48\%$). 

The significant detections we found for this wavelength are comparable to those of large, optically selected AGN samples in the local universe (e.g., \(\sim 80-90\%\) AGN purity in SDSS BPT-selected samples, \citealt{kauffmann03, kewley06}). This high detection rate is expected given that the optical spectroscopic selection traces emission from the NLR, similar to the [Ne\,{\sc v}] selection. Therefore, AGN with developed NLRs as our sample should be naturally detectable by both methods. However, it is significantly higher than the \(41.2 \pm 8.9\%\) reported by \citetalias{goulding09} for their [Ne\,{\sc v}] sample, which contained a higher proportion of LINERs and H\,{\sc ii} regions. This discrepancy likely arises from differences in the sample selection and classification method in which \citetalias{goulding09} applied uniform optical diagnostics to a heterogeneous sample, while our classifications are compiled from NED, which aggregates results from multiple studies and may include updated reclassifications.

To investigate whether the optically unidentified AGN in our sample is affected by obscuration due to the inclination of the host galaxy, we examined the distribution of the galaxy inclination angles of our sources as shown in Figure~\ref{fig:inclination}. The plot reveals no significant correlation between optical detection and galaxy orientation, in which a Kolmogorov-Smirnov (KS) test shows no statistical difference between the inclination distributions of optical and non-optical AGN (\(D = 0.22, P_{\text{KS}} = 0.34\)). Mean inclinations are similar for both subsets (\(56.7^\circ \pm 18.3\) for optical vs. \(62.0^\circ \pm 18.8\) for non-optical AGN), and both span the full range from face-on to edge-on systems. This indicates that the galaxy inclination angle or host galaxy obscuration does not affect our optical AGN selection rate. In addition, we also performed visual inspection on the galaxy images to investigate whether there are prominent merger or tidal features that may affect optically identified and unidentified AGN. However, we did not find any significant differences between the morphologies of both groups of galaxies that host optically identified and unidentified AGN. This further suggests that the obscuration does not affect the detection rate. In contrast, this result is not consistent with the findings of \citetalias{goulding09}, who reported that their optically unidentified [Ne\,{\sc v}] AGN (10/17, $58.82 \pm 8.92\%$) missed, either reside in highly inclined galaxies or those with dust lanes, obscuring the AGN signatures.

We also investigated whether low AGN luminosity might be the reason for non-detections. However, we found that all optically unidentified sources in our sample have luminosities comparable to those identified with the optical selection method. Therefore, other factors may play a role in non-detection such as high star formation activity that may dilute optical emissions from the nucleus \citep{trump15, lambrides20}. This is supported by Figure~\ref{fig:wise}, where all (except one) of the optically undetected sources lie outside the AGN wedge, consistent with a dominant contribution from star-forming activity. In addition, most of them occupy bluer \(W1-W2\) and redder \(W2-W3\) regions of the diagram compared to the optically detected AGN. This indicates stronger star-forming contributions in the optically unidentified AGN sources.

\subsection{Summary}
\label{subsec:multiwavelength_compare}

\begin{figure}
\centering
	\includegraphics[width=0.7\columnwidth]{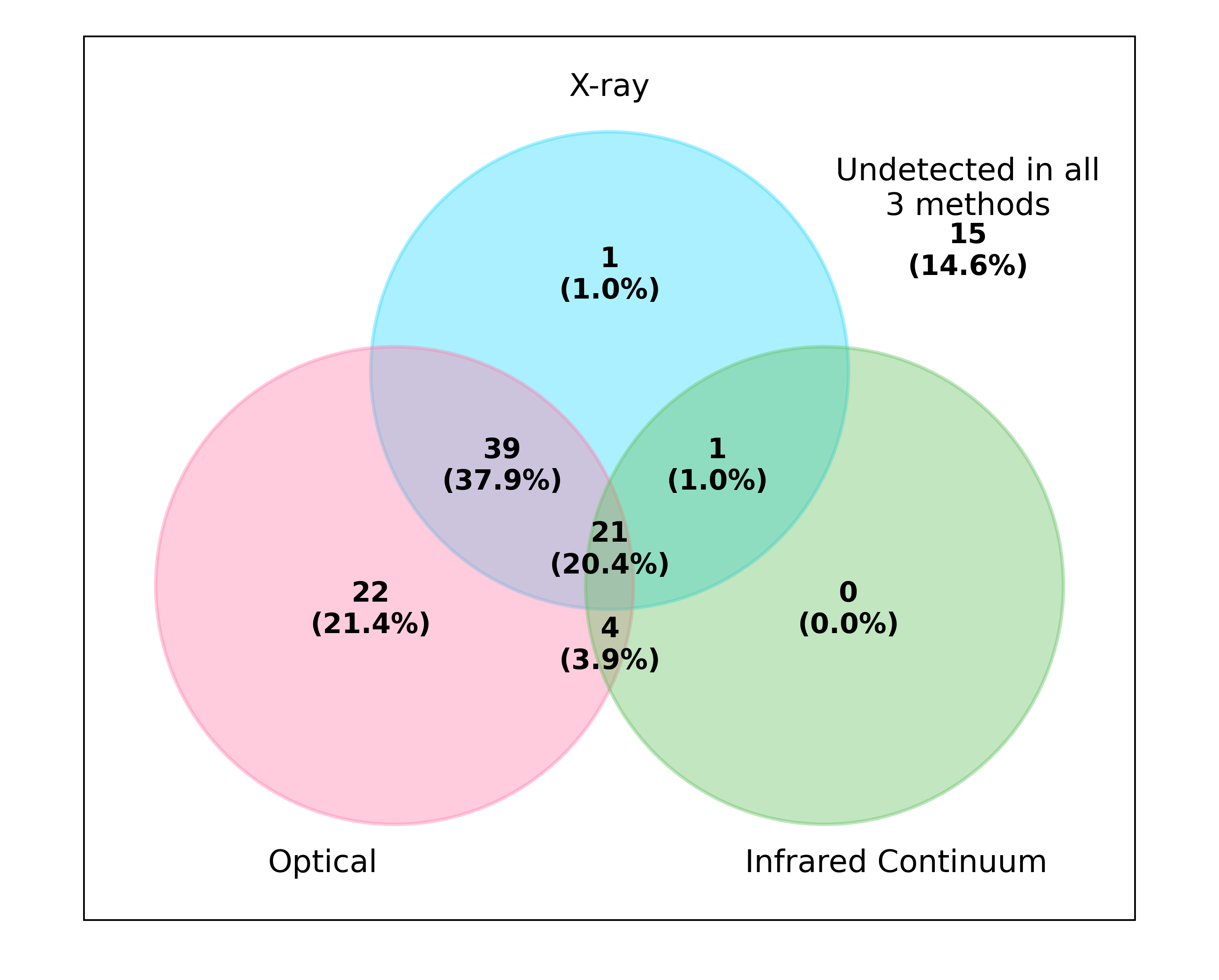}
    \caption{Venn diagram of multiwavelength overlap in AGN selection for our [Ne\,{\sc v}]-selected sample. The diagram reveals various value of incompleteness in all traditional selection methods, with optical diagnostics performing best (83.6\% recovery), hard X-ray (60\% recovery) and mid-infrared colors performing poorest (25\% recovery). Most notably, 15 sources (14.6 \%) remain undetected by all three classical methods, demonstrating the unique capability of [Ne\,{\sc v}] selection to identify AGN missed by traditional surveys.}
    \label{fig:venn-agn_dis}
\end{figure}




\begin{deluxetable}{lcc}




\tablecaption{Multiwavelength AGN detection rates of the complete AGN sample.}
\label{detection_rates}

\tablenum{1}

\tablehead{\colhead{\textbf{Wavelength}} & \colhead{\textbf{N}} & \colhead{\textbf{Detection rate (\%)}} \\ 
\colhead{(1)} & \colhead{(2)} & \colhead{(3)} \\} 

\startdata
X-ray & 103 & 60.19 $\pm$ 8.87 \\
Optical & 103 & 83.50 $\pm$ 6.73 \\
Optical (based on data availability) & 96 & 89.58 $\pm$ 6.11 \\
Infrared (continuum) & 103 & 25.24 $\pm$ 7.87 \\
Infrared (continuum) (based on data availability) & 88 & 29.55 $\pm$ 9.53 \\
\enddata


\tablecomments{(1) Wavelength band used. (2) The total number of sample considered in calculation for each subset. (3) Detection rates accounting to the full volume-limited [Ne v] sample ($N=103$) and also to sources with available data in the corresponding band.}

\end{deluxetable}

We have compared our AGN sample with three traditional multiwavelength AGN selection methods to understand the demographics and properties of our sample. Optical diagnostics (using NED classifications) recovered the highest fraction at $89.6 \pm 6.1\%$, followed by hard X-ray selection (\textit{Swift}/BAT) at \(60.2 \pm 8.9\%\) and finally mid-infrared color selection (\textit{WISE}) at $29.6 \pm 9.5\%$. The Venn diagram (Figure~\ref{fig:venn-agn_dis}) visually consolidates these multiwavelength detection patterns. The high optical recovery is consistent with the fact that our sample is biased towards AGN with well-developed NLR. The small fraction not detected may be due to the high star formation activity in the galaxy that dilutes the optical signature from the nucleus. Moderate X-ray recovery is in agreement with previous studies (\citealt{goulding10}; \citealt{ricci15, ananna19}) and suggests a significant fraction of heavily Compton-thick AGN. The low MIR continuum recovery points to a widespread host-galaxy dilution that evades broadband photometric cuts \citep{asmus20, lambrides20}.

Figure~\ref{fig:venn-agn_dis} also highlights both the complementarity and limitations of different detection methods. AGN detected only in X-rays constitute a negligible fraction ($1.0 \pm 1.4\%$) of the sample, reflecting the rarity of systems whose accretion signatures are exclusively visible above the X-ray detection threshold without optical or infrared counterparts. In contrast, X-ray and optical overlap is dominant ($37.9 \pm 8.8\%$), indicating that the majority of AGN selected by X-ray are also optically identifiable, strengthening the consistency between these two common selection techniques. Optical-only detections make up a substantial portion ($21.4 \pm 7.4\%$), likely representing low-luminosity or host-dominated AGN whose faint X-ray emission falls below the sensitivity of \textit{Swift}/BAT. In particular, no AGN ($0.0\%$) are detected exclusively in the infrared continuum, suggesting that in this sample selected by [Ne\,{\sc v}], mid-infrared AGN signatures rarely occur without accompanying X-ray or optical emission. Infrared-optical overlap ($3.9 \pm 2.8\%$) constitutes a small population where the AGN is detectable in reprocessed dust emission and optical lines, but weak in X-rays. Notably, no sources in our sample were detected exclusively in X-ray and infrared without an optical counterpart, implying that heavily obscured AGN residing in strong star forming galaxies often predicted in galaxy evolution models are rare or absent in our [Ne\,{\sc v}]-selected sample. Systems detected across all three bands ($20.4 \pm 7.3\%$) correspond to luminous, unobscured AGN with clear multi-wavelength signatures.

The most significant finding in Figure~\ref{fig:venn-agn_dis} is the “[Ne\,{\sc v}]-only” population, in which 14.6 ± 6.4\% (15/103) of our sources evade detection by all three standard methods and are only identified by the [Ne\,{\sc v}] line. This population may represent a critical gap in current AGN surveys that likely includes heavily Compton-thick and/or host-dominated systems that are missed by classical optical, X-ray, and IR selection. Our results demonstrate that even the combination of standard multi-wavelength methods is incomplete and that [Ne\,{\sc v}] spectroscopy provides an independent detection channel to uncover the full AGN population.

\section{AGN Properties}
\label{sec:agn-properties}

In this section, we investigated the fundamental AGN properties of our [Ne\,{\sc v}]-selected sample, focusing on bolometric luminosities, black hole masses, and Eddington ratios to characterize this population. We present Table~\ref{appA} for the main result of AGN properties for each source.

\subsection{Bolometric Luminosity}
\label{subsec:Lbol}

Bolometric luminosity represents the total energy output of an AGN and is crucial to understanding accretion physics and energetic feedback. We derived $L_{{\rm bol}}$ from [Ne\,{\sc v}] luminosities to characterize the power output of our sample. We estimated the bolometric luminosities using a correction of $L_{{\rm bol}} = \kappa \times L_{\mathrm{[Ne\,v]}}$ where $\kappa = 13000$ (\citetalias{dasyra08}). This correction factor was originally calibrated against 5100\,\AA\, continuum luminosities, and benefits from the direct dependence of narrow-line region emission, including [Ne\,{\sc v}], on the AGN accretion rate. To validate this approach, we compared our [Ne\,{\sc v}]-derived $L_{{\rm bol}}$ with the available archival SED-fitting $L_{{\rm bol}}$. Of 103, only 67 have SED-fitted bolometric luminosities. We collected $L_{{\rm bol}}$ of 67 sources from \citet{ richards06-sed, dopita14-sed, barrows21-sed, gupta24-sed, gianolli24-sed}. We constructed a regression line fit between SED-derived bolometric luminosities and estimates based on [Ne\,{\sc v}]. Figure~\ref{fig:fit} shows the relationship between the two parameters with the best-fit regression line. The results of the fitting are shown in Table~\ref{fitting}. The best-fit solution is given by

\begin{equation} \label{eq:1}
    \left( \frac{\log\,L_{\rm bol\,\rm[Ne\, v]}}{\rm erg/s} \right) = (8.83 \pm 3.40) + (0.82 \pm 0.08) \left( \frac{\log\,L_{\rm bol\,\rm SED}}{\rm erg/s}\right)
\end{equation}

\noindent with parameter values $\alpha = 0.82 \pm 0.08$ and $\beta = 8.83 \pm 3.40$. The fit reveals a strong correlation (Pearson's $r = 0.8$) between [Ne\,{\sc v}]-based $L_{{\rm bol}}$ and SED-based $L_{{\rm bol}}$, with a moderate residual scatter of 0.6 dex. This provides evidence for the reliability of our data. We therefore proceeded to use the [Ne\,{\sc v}]-based $L_{{\rm bol}}$ for the whole sample for consistency.

We measured a bolometric luminosity range of \(3.8 \times 10^{41}\) to \(4.9 \times 10^{45} \,\mathrm{erg\,s^{-1}}\) for our [Ne\,{\sc v}] sample, with a mean of \(\log(L_{\mathrm{bol}}/\mathrm{erg\,s^{-1}}) = 44.5 \pm 0.7\). This mean luminosity is consistent with that of the hard X-ray-selected AGN sample from \textit{Swift}/BAT (\(\log L_{\mathrm{bol}} = 44.4 \pm 0.7\); \citealt{koss17}). Figure~\ref{fig:agn-properties} shows the comparison between the luminosity distributions of our sample and the BAT sample, showing that the two distributions are all in good agreement. A two-sample Kolmogorov–Smirnov (KS) test confirms this similarity (\(D_{\rm KS} = 0.13, P_{\rm KS} \sim 0.21\)). The luminosity consistency between the samples, despite fundamentally different selection methods, indicates that both the [Ne\,{\sc v}] emission and hard X-rays trace AGN populations with comparable energetic output. This suggests that [Ne\,{\sc v}] selection does not introduce strong luminosity biases relative to established X-ray methods. The range of our derived luminosity is also consistent with previous [Ne\,{\sc v}] studies, which report \(\log L_{\mathrm{bol}}\) values spanning \(\sim\)40–45 (e.g., \citetalias{goulding09, annuar25}). The agreement among independent studies reinforces the reliability of the [Ne\,{\sc v}] technique and suggests that it consistently identifies AGN across a characteristic luminosity range that bridges low-luminosity Seyferts and the more powerful quasars.

\begin{deluxetable}{lcccccc}
\tablecaption{Best-fit parameters for $M_{\rm BH}$ vs. $L_{\rm [Ne\,v]}$}
\label{fitting}
\tablenum{2}

\tablehead{
\colhead{Subsample} &
\colhead{$N$} &
\colhead{Slope ($\alpha$)} &
\colhead{Intercept ($\beta$)} &
\colhead{$\sigma$ (dex)} &
\colhead{$r$} &
\colhead{$p$-value}
}

\startdata
RM & 23 & $0.81 \pm 0.15$ & $7.86 \pm 0.10$ & 0.45 & 0.76 & $2.2\times10^{-5}$ \\
$M_{\rm BH}$--$\sigma_\ast$ & 41 & $0.61 \pm 0.11$ & $7.58 \pm 0.07$ & 0.39 & 0.66 & $3.9\times10^{-1}$ \\
Combined archival & 77 & $0.42 \pm 0.10$ & $7.73 \pm 0.07$ & 0.59 & 0.43 & $7.8\times10^{-5}$ \\
\enddata

\tablecomments{Best-fit parameters for the relations between black hole mass $M_{\rm BH}$ and $[\mathrm{Ne\,v}]$ 14.32 $\mu$m luminosity $L_{\rm [Ne\,v]}$ based on available established masses for our sample ($N$). We performed line fit for 3 groups of black hole mass methods mainly the reverberation mapping (RM), then the $M_{\rm BH}-\sigma_{\ast}$ relation method, and lastly is combined methods including RM, direct dynamical measurements (stellar and gas dynamics, and megamasers) and host galaxy scaling relation ($M_{\rm BH}-\sigma_{\ast}$). We list out the slope, intercept, intrinsic scatter, Pearson correlation coefficient ($r$), and $p$-value for all three fits. We found that the RM method subsample shows the strongest correlation ($r \approx 0.76$) with moderate scatter ($\sim0.4$ dex) therefore it is adopted in this work to estimate $M_{\rm BH}$ for the remaining sample with no available established $M_{\rm BH}$.}

\end{deluxetable}

\begin{figure*}
    \centering
	\includegraphics[width=0.45\textwidth]{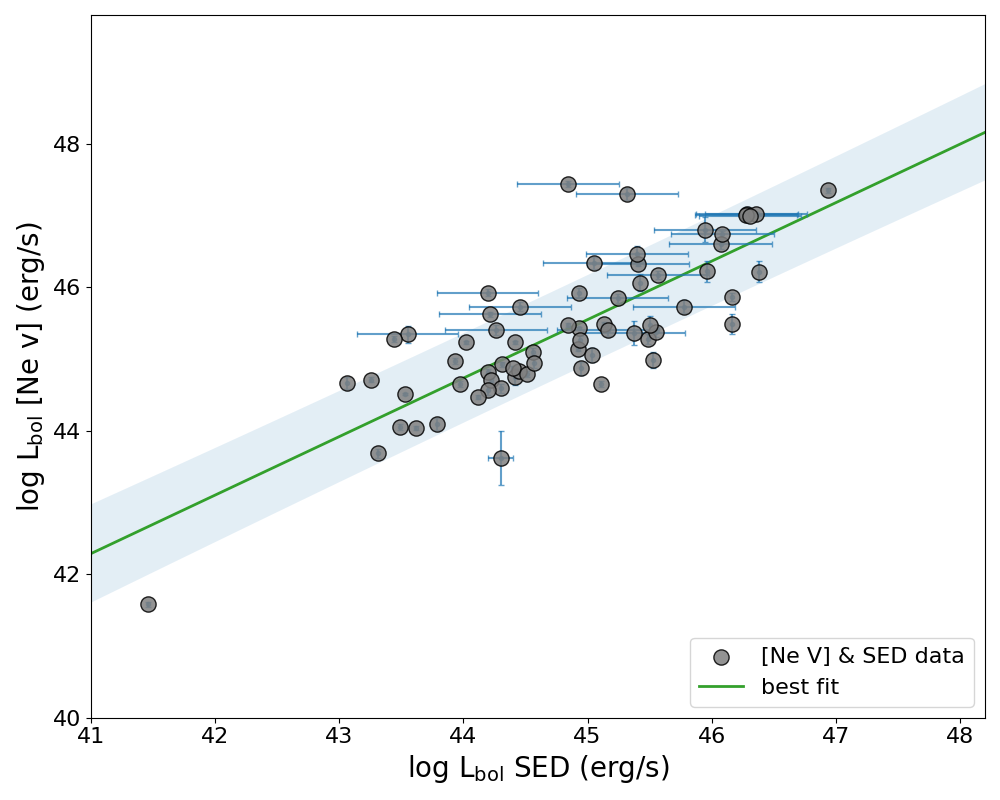}
    \includegraphics[width=0.45\textwidth]{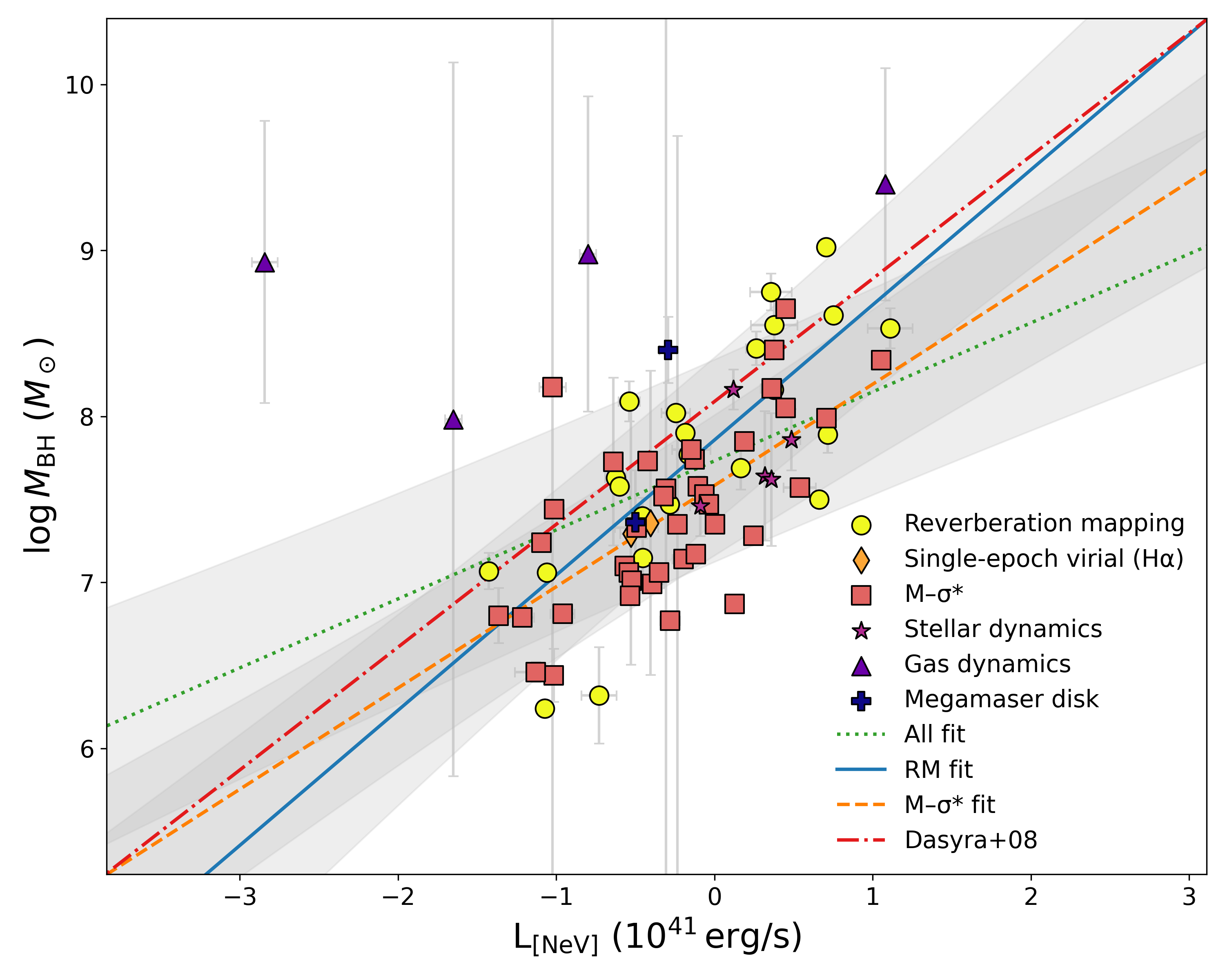} \\
    \caption{\textit{Left}: Comparison between bolometric luminosities derived from [Ne\,{\sc v}] emission and archival SED fitting for 67 sources. The best-fit relation (solid line) and its uncertainty (shaded region) show a strong correlation ($r = 0.8$) with a scatter of 0.6 dex, indicating that [Ne\,{\sc v}] is a reliable tracer of bolometric luminosity. \textit{Right}: Calibration of the $M_{\mathrm{BH}}$–$L_{\mathrm{[Ne\,V]}}$ relation using archival black hole masses from reverberation mapping, host-galaxy scaling relations, and direct dynamical methods. Best-fit relations are shown for the full sample, the RM subsample, and the $M_{\mathrm{BH}}$–$\sigma_*$ subsample, together with the relation from \citetalias{dasyra08}. The RM subsample shows the strongest correlation ($r \approx 0.8$) and is adopted in this work.}
    \label{fig:fit}
\end{figure*}

\subsection{Black Hole Mass}
\label{subsec:mbh}

\begin{figure*}
    \centering
    \includegraphics[width=0.45\textwidth]{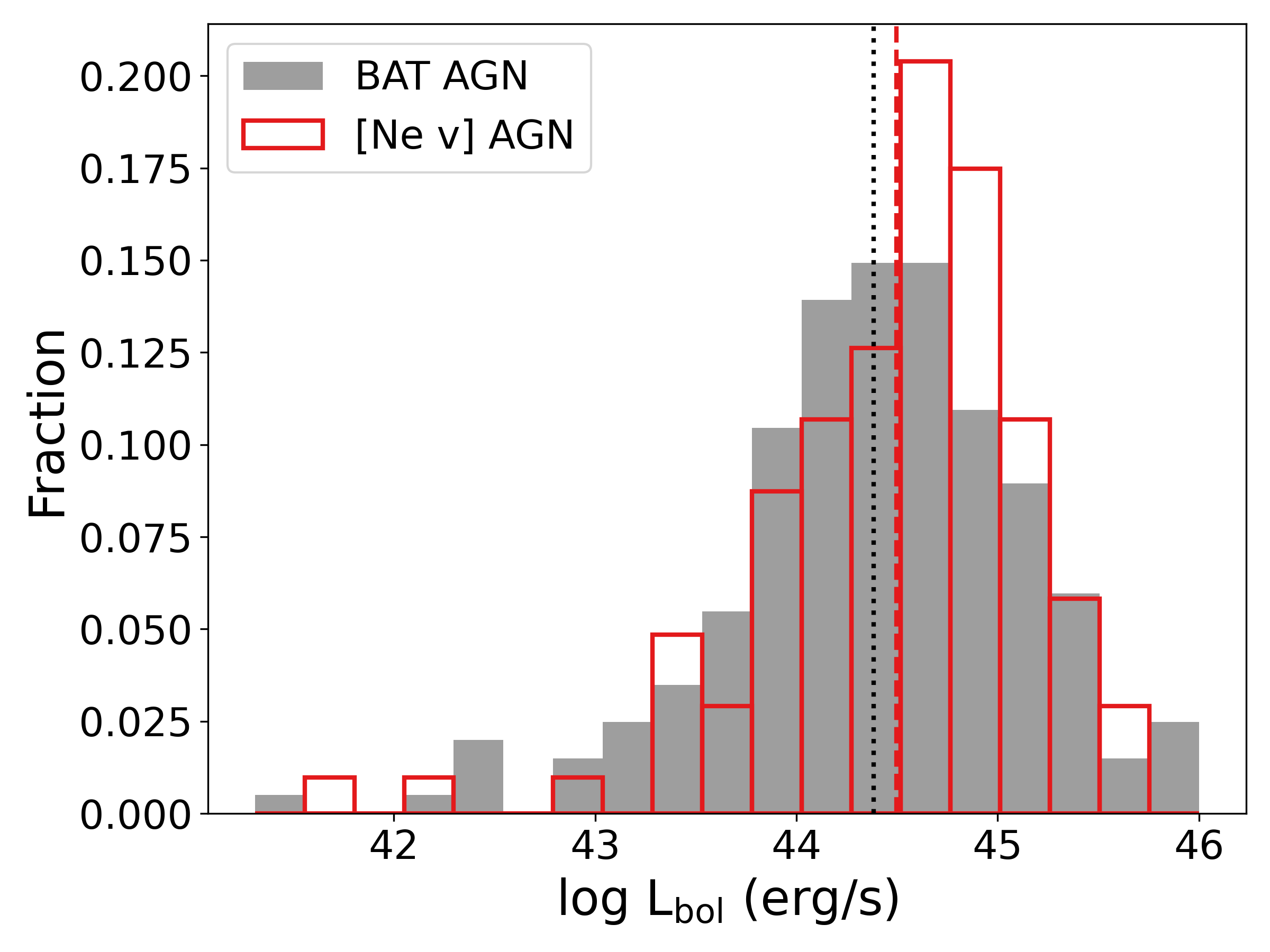}\\
    \includegraphics[width=0.45\textwidth]{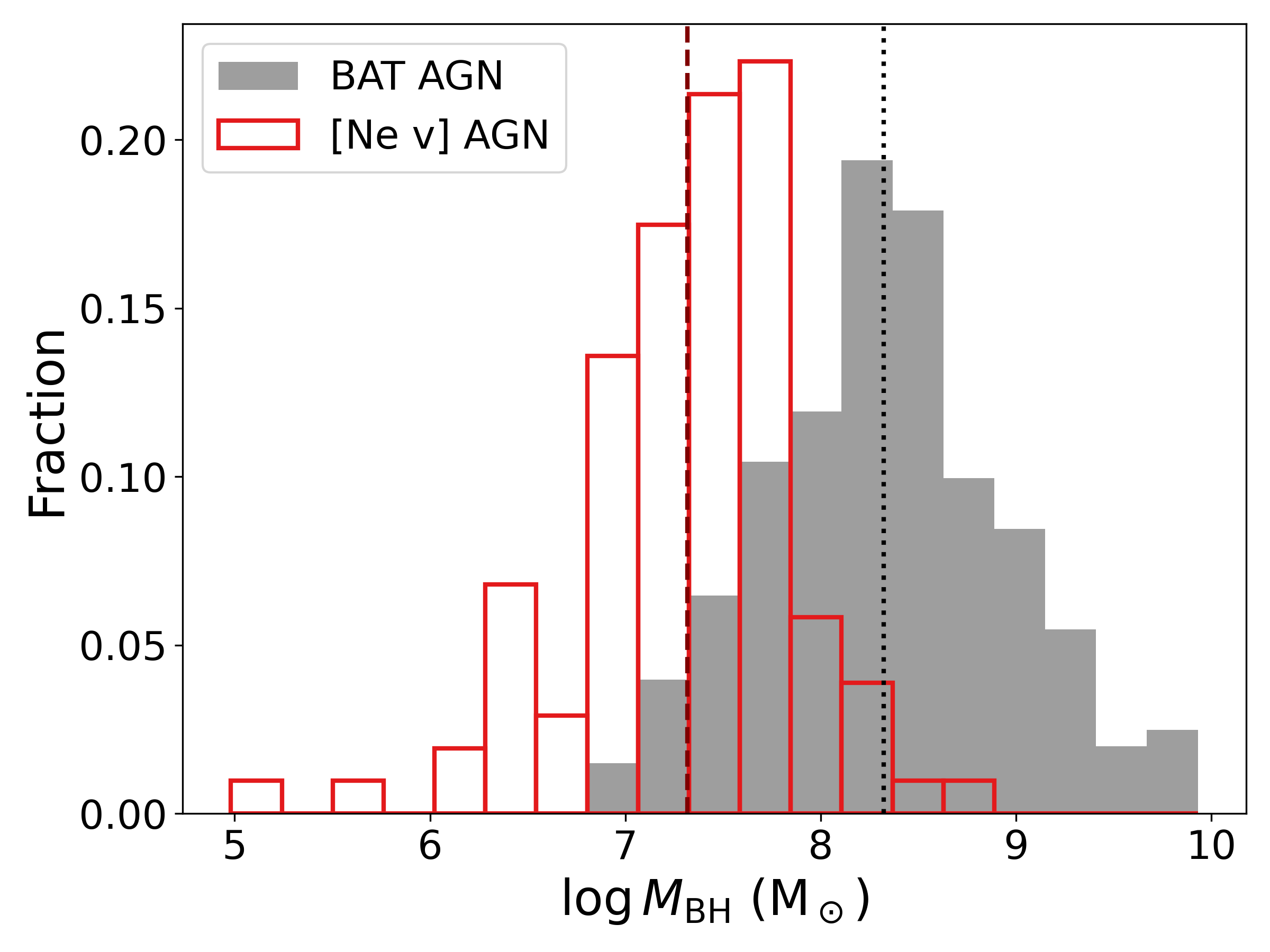}
    \includegraphics[width=0.45\textwidth]{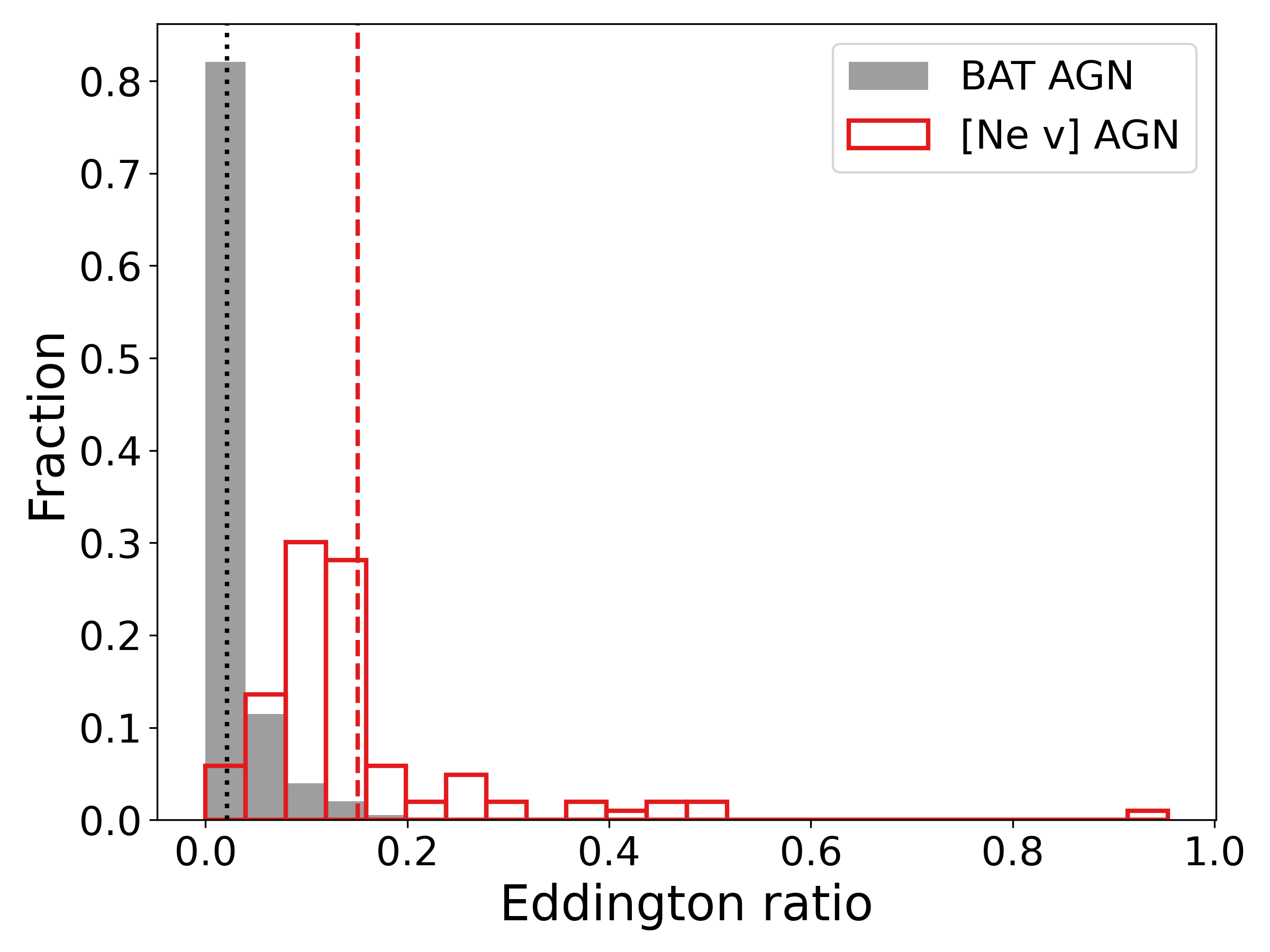} \\
    \caption{Distribution of AGN physical properties for the complete [Ne\,{\sc v}]-selected sample (red line) compared to the Swift/BAT sample from \citet{koss17} (gray fill). Colored vertical dashed and dotted lines denote the statistical mean of each distribution. Parameter distributions showing bolometric luminosity ($L_{\rm bol}$; \textit{left}), black hole mass ($M_{\rm BH}$; \textit{center}), and Eddington ratio $\lambda_{\rm Edd} = L/L_{\rm Edd}$ (\textit{right}). The [Ne\,{\sc v}] sample exhibits consistent luminosity, lower black hole masses, and higher Eddington ratios compared to the X-ray selected \textit{Swift}/BAT sample.}
    \label{fig:agn-properties}
\end{figure*}

The black hole mass ($M_{\rm BH}$) is a fundamental parameter governing the energetics and evolution of AGN. Obtaining black hole mass is important in determining the accretion power of AGN, which is vital for producing the distribution of AGN accretion evolution over time. We obtained our $M_{\rm BH}$ for our sources using established $M_{\rm BH}$, otherwise we estimated the masses using the [Ne\,{\sc v}] line. We investigated the relationship between $M_{\rm BH}$ and the luminosity of [Ne\,{\sc v}] (\citetalias{dasyra08}). We derived best fit lines based on various established methods such as reverberation mapping, single-epoch virial, stellar and gas kinematics, megamasers, and the host-galaxy scaling relation, the $M_{{\rm BH}}$–$\sigma^*$ relation.

We tested for relations using three subsamples of black hole mass measurements. The first subsample consists of masses obtained from broad-line region (BLR)–based methods, specifically reverberation mapping (RM). The second subsample comprised indirect measurements based on host-galaxy, using the $M_{{\rm BH}}-\sigma^*$ relation. The third subsample included archival measurements using all methods, combining BLR-based masses (RM and single virial), direct dynamical estimates (stellar dynamics, gas kinematics and megamaser disks), and indirect host-galaxy scaling relations. We focused primarily on the RM and $M_{{\rm BH}}-\sigma^*$ subsamples because they represented the most reliable direct AGN-based estimators (RM) and the most widely available indirect (host-based) estimators, allowing us to test the relation [Ne\,{\sc v}]–$M_{\mathrm{BH}}$ between fundamentally different calibration methods while minimizing systematic uncertainties from heterogeneous measurement techniques.

Table~\ref{fitting} summarizes the best-fit parameters for the three subsamples. Figure~\ref{fig:fit} presents the corresponding best-fit relations, together with the relation $M_{\text{BH}}-L_{\rm [Ne \,v]}$ from \citetalias{dasyra08}. Among the three subsamples, the RM sample exhibits the strongest correlation (Pearson $r = 0.76, p=2.2\times10^{-5}$), compared to the $M_{{\rm BH}}-\sigma^*$ subsample ($r = 0.66, p=3.9\times10^{-1}$) and the combined measurements ($r = 0.43, p=7.8\times10^{-5}$). Given the tighter correlation and the more direct physical connection to the active nucleus, we adopted the RM-based best-fit relation to estimate $M_{{\rm BH}}$ for sources lacking archival mass measurements. We present the best-fit solution for the RM subsample as follows:

\begin{equation} \label{eq:2}
    \left(\frac{\log\,M_{\rm BH}}{\rm M_\odot}\right) = (7.86 \pm 0.10) + (0.81 \pm 0.15) \left(\frac{\log\,L_{\rm [Ne \,v]}}{\rm 10^{41} erg/s}\right) 
\end{equation}

\noindent with parameter values corresponding to $\alpha = 0.81 \pm 0.15$ and $\beta = 7.86 \pm 0.10$. The scatter found from this correlation ($\sim$0.45 dex) is consistent with that determined by \citetalias{dasyra08} (0.46 dex). The scatter is also remarkably similar to other single-epoch mass estimators \citep{greene06, ho_kim15, bennert21}. This suggests that [Ne\,{\sc v}] provides precision comparable to established methods, while offering the unique advantage of working in heavily obscured systems where traditional diagnostics fail \citep{comastri15, onori17, lamperti17}.

Figure~\ref{fig:agn-properties} shows the $M_{{\rm BH}}$ distribution of our sample. Our sample has a black hole mass range of $9.6 \times 10^4 - 4.5 \times 10^8 M_\odot$, median and mean log black hole mass of $7.4 \pm 0.6$ and $7.3 \pm 0.6 ,M_\odot$, respectively. The lowest mass we found is of NGC 4395, which is a dwarf galaxy \citep{mezcua24}. 

Comparing our sample with the \textit{Swift}/BAT, we found that our sources have a lower maximum $M_{{\rm BH}}$ (approximately 1 order of magnitude) than the BAT sample and span a region of low $M_{{\rm BH}}$ masses that the hard X-ray survey does not cover. Our mean $M_{{\rm BH}}$ is also 1 order of magnitude lower than BAT (mean $\log M_{{\rm BH}} = 8.3 \pm 0.6 \,M_\odot$). These results are generally consistent with those found by \citetalias{annuar25}, which also determined lower $M_{{\rm BH}}$ masses for their [Ne\,{\sc v}] sample compared to \textit{Swift}/BAT. Although the mean values between our AGN and the BAT survey are consistent within the uncertainties, the KS test reveals fundamentally different distribution shapes between the two samples ($D_{\rm KS} =0.62$, $P_{\rm KS} \sim9.9 \times 10^{-28}$). The systematic offset toward lower masses in our [Ne\,{\sc v}]-selected sample suggests that this method preferentially identifies lower mass black holes compared to hard X-ray selection and covers a population of low mass black holes not detected by \citet{koss17}. Alternatively, \citet{koss17} selection is similar to \citet{caccianiga07}'s Type 1 AGN masses and may be biased toward more massive systems or more archetypal AGN masses. 

\subsection{Eddington Ratio}
\label{subsec:edd}

The Eddington ratio ($\lambda_{\mathrm{Edd}} = L_{{\rm bol}}/L_{\mathrm{Edd}}$) quantifies accretion efficiency relative to the theoretical maximum, providing information on the fueling and growth phases of AGN. We examined the accretion characteristics of our [Ne\,{\sc v}]-selected sample to understand their growth states and evolutionary implications. We calculated the Eddington ratios using our bolometric luminosity determinations (see Section~\ref{subsec:Lbol}), and  $M_{\rm BH}$ from the literature if available otherwise, we estimated the masses using their [Ne\,{\sc v}] luminosities (see Section~\ref{subsec:mbh}). This approach ensures comprehensive coverage across our sample.

As can be seen in Figure~\ref{fig:agn-properties}, the distribution of the Eddington ratio shows a range of $\lambda_{\rm Edd} = 0.03 - 0.95$, and a mean $\lambda_{\rm Edd} = 0.15 \pm 0.1$. The sample represents a wide distribution of AGN accretion from low to near-Eddington processes; and on average, is more frequent in the efficient and higher accretion power region. Figure~\ref{fig:agn-properties} also reveals a distribution with a long tail up to $\lambda_{\mathrm{Edd}} \sim 0.5$ and an extreme source at $\lambda_{\mathrm{Edd}} \sim 1$. The extended high-accretion tail demonstrates that our sample includes a large number of systems undergoing particularly rapid growth phases.

Compared to the \textit{Swift}/BAT AGN \citep{koss17}, the mean of BAT is $\lambda_{\mathrm{Edd}} = 0.02 \pm 0.03$, significantly lower than our sample. The KS test shows highly significant differences ($D_{\rm KS} =0.76$, $P_{\rm KS} \sim1.5 \times 10^{-39}$) between the two samples, with \citet{koss17} showing a rapidly declining distribution concentrated at lower accretion rates ($\lambda_{\mathrm{Edd}} < 0.01$). We observed a wider range and higher values of the Eddington ratio in our sample compared to the BAT sample, which can also be seen without considering the outlier at $\rm \lambda_{\mathrm{Edd}} \sim 1$. The outlier galaxy with $\lambda_{\mathrm{Edd}}= 0.95$, UGC 01214, is a known Seyfert 2 \citep{springob05} that has been found to be radiating near or above the Eddington limit \citep{kraemer09}.

\section{Summary and Conclusions}
\label{conclusions}

Our understanding of the cosmic AGN population has long been reliant on independent selection techniques, each with their own advantages and selection biases The high-ionization mid-infrared [Ne\,{\sc v}] emission line serves as a powerful AGN-specific tracer that is not significantly affected by dust obscuration and contamination. Therefore, it can act as a significant method in identifying AGN to complement more classical techniques. This study used a volume-limited ($z\leq0.025$) [Ne\,{\sc v}]-selected sample from 103 AGN to evaluate the properties and demographics of our sample through multiwavelength analysis. We found that:

\begin{enumerate}[label=(\roman*)]
    \item our [Ne\,{\sc v}]-selected sample missed $\sim$45\% of the local AGN population identified by hard X-ray and mid-infrared continuum methods (NLR-independent methods) indicating that [Ne\,{\sc v}] selection is biased against systems with under-developed NLRs;
    
    \item hard X-ray selection (\textit{Swift}/BAT) identified $60.2 \pm 8.9\%$ counterparts for our AGN suggesting that $\sim 40\%$ of the AGN, was missed using this method. Based on our analysis we found that out of the 40\% X-ray unidentified AGN, 10\% are genuinely not detected due to \textit{Swift}/BAT sensitivity limit, while the remaining 30\% are likely to be hidden by heavy Compton-thick obscuration with ${N_{\mathrm{H}} \gtrsim 10^{25}}$ cm$^{-2}$ that remain buried even in the hard X-ray band;
    
    \item optical selection missed $\sim10\%$ of our [Ne\,{\sc v}] sample, while mid-infrared color selection missed $\sim70\%$ of the population. In both cases, we determined that the primary cause is dilution by host-galaxy emission;
    
    \item interestingly, $14.6 \pm 6.4\%$ of the [Ne\,{\sc v}] AGN were not detected by any of the other selection methods. This “[Ne\,{\sc v}]-only” population demonstrates the line's role as a complementary tool for identifying a complete AGN population.  
\end{enumerate}  

We also derived physical properties for our sample, namely bolometric luminosity, black hole mass, and accretion rate to investigate the AGN demographics and compare them with the \textit{Swift}/BAT AGN sample. We determined that:

\begin{enumerate}[label=(\roman*)]
    \item the sample has a bolometric luminosities range of $10^{41} - 10^{45} \,\mathrm{erg\,s^{-1}}$, and mean \(\log(L_{\rm bol}/\rm erg\,s^{-1}) = 44.65 \pm 0.47\). This is consistent with the typical AGN sample from the \textit{Swift}/BAT survey; 
    \item our sample has an $M_{\rm BH}$ range of $10^4 - 10^8 M_\odot$ and mean of \(\log(M_{\rm BH}/M_{\odot}) = 7.49 \pm 0.43\), systematically lower than the X-ray selected sample; and
    \item the Eddington ratios for our sample; i.e., $\lambda_{\mathrm{Edd}} = 0.15 \pm 0.11$, are significantly higher than the \textit{Swift}/BAT survey. Our sample is characterized by a wide accretion distribution of $0.03$–$1$ indicating that the [Ne\,{\sc v}] line selects both low and high accretion powered sources.
\end{enumerate}

The comparable bolometric luminosities, but lower black hole masses and correspondingly higher Eddington ratios of our [Ne\,{\sc v}] sample relative to the \textit{Swift}/BAT survey are reminiscent of the rapidly accreting AGN population identified by \citet{aird12}, which was found to peak at cosmic noon. This suggests that our [Ne\,{\sc v}] selected AGN may represent local analogs of these efficiently accreting systems. This adds to the importance of utilizing [Ne\,{\sc v}] as an AGN tracer. Future observations with the, e.g., \textit{James Webb Space Telescope} can extend the [Ne\,{\sc v}] studies of AGN to higher redshifts, connecting these local efficiently accreting systems to cosmic AGN evolution and enabling a more complete census of black hole growth across time.

\begin{acknowledgments}
The authors thank the referee for their constructive and thorough feedback, which has significantly improved the article. The authors also thank D. M. Alexander for useful discussion of this work. We acknowledge financial support from Universiti Kebangsaan Malaysia's Faculty of Science and Technology internal grant (Research BOOST Grant FST1-PNI). This research has used the NASA/IPAC Extragalactic Database (NED), which is funded by the National Aeronautics and Space Administration and operated by the California Institute of Technology. This work is mainly based on observations made with the Spitzer Space Telescope, which was operated by the Jet Propulsion Laboratory, California Institute of Technology, under a contract with NASA. This research has used the VizieR catalog access tool, CDS, Strasbourg, France (\doi{10.26093/CDS/VIZIER}). The original description of the VizieR service was published in 2000, $\rm A\&AS$ 143, 23.
\end{acknowledgments}

\appendix

\begin{longrotatetable}
\section{Complete list of [Ne v]-selected AGN at $z \leq 0.025$}
\begin{deluxetable}{lcccccccccccc}



\tablecaption{Properties of the complete [Ne\,{\sc v}]-selected AGN sample at $z\leq0.025$.}
\label{appA}

\tablenum{3}

\tablehead{\colhead{\textbf{AGN}} & \multicolumn{2}{c}{\textbf{Position}} & \colhead{\textbf{Redshift}} & \multicolumn{3}{c}{\textbf{Detection}} & \multicolumn{2}{c}{\textbf{Luminosities}} &  \multicolumn{3}{c}{\textbf{Black Hole Mass Properties}} & \colhead{$\boldsymbol{\lambda}_{{\text{Edd}}}$} \\
 & \colhead{RA} & \colhead{Dec} &  & \colhead{X-ray} & \colhead{Optical} & \colhead{\shortstack{Infrared\\colors}} & \colhead{$\log L_{\rm [Ne \,v]}$} & \colhead{$\log L_{\rm bol}$} & \colhead{$\log M_{\rm BH}$} & \colhead{Method} & \colhead{Ref.} &  \\ 
\colhead{} & \colhead{(deg)} & \colhead{(deg)} & \colhead{} & \colhead{} & \colhead{} & \colhead{} & \colhead{(erg\,s$^{-1}$)} & \colhead{(erg\,s$^{-1}$)} & \colhead{(M$_\odot$)} & \colhead{} & \colhead{} & \colhead{} } 

\startdata
\shortstack{2MASX J03250535\\+4033310} & 51.27 & 40.56 & 0.023 & $\times$ & Unclassified & ... & $40.75 \pm 0.10$ & $44.86 \pm 0.10$ & $7.65^{+0.11}_{-0.11}$ & [Ne v] & This work & $0.13^{+0.04}_{-0.04}$ \\
\shortstack{2MASX J04344151\\+4014219} & 68.67 & 40.24 & 0.02 & $\times$ & Sy 1 & $\times$ & $40.91 \pm 0.07$ & $45.02 \pm 0.07$ & $7.78^{+0.10}_{-0.10}$ & [Ne v] & This work & $0.14^{+0.04}_{-0.04}$ \\
\shortstack{2MASX J06594021\\-6317531} & 104.92 & -63.3 & 0.023 & $\times$ & H\,{\sc ii} & $\times$ & $40.42 \pm 0.09$ & $44.53 \pm 0.09$ & $7.39^{+0.13}_{-0.13}$ & [Ne v] & This work & $0.11^{+0.04}_{-0.04}$ \\
\shortstack{2MASX J13463217\\+6423247} & 206.63 & 64.39 & 0.024 & $\times$ & Sy 1 & $\times$ & $40.04 \pm 0.08$ & $44.16 \pm 0.08$ & $6.81^{+0.00}_{-0.00}$ & M-$\sigma^*$ & (1) & $0.18^{+0.03}_{-0.03}$ \\
\shortstack{2MASX J16191179\\-0754026} & 244.8 & -7.9 & 0.024 & $\times$ & LINER & ... & $40.75 \pm 0.06$ & $44.86 \pm 0.06$ & $7.65^{+0.11}_{-0.11}$ & [Ne v] & This work & $0.13^{+0.04}_{-0.04}$ \\
CGCG 141-034 & 269.24 & 24.02 & 0.02 & $\times$ & Unclassified & $\times$ & $40.53 \pm 0.06$ & $44.64 \pm 0.06$ & $7.47^{+0.12}_{-0.12}$ & [Ne v] & This work & $0.12^{+0.04}_{-0.04}$ \\
CGCG 142-033 & 274.14 & 22.11 & 0.018 & $\times$ & Unclassified & $\times$ & $40.16 \pm 0.08$ & $44.27 \pm 0.08$ & $7.17^{+0.16}_{-0.16}$ & [Ne v] & This work & $0.10^{+0.04}_{-0.04}$ \\
\shortstack{CGCG 468-002\\NED01} & 77.08 & 17.36 & 0.018 & / & Sy 2 & $\times$ & $40.32 \pm 0.05$ & $44.44 \pm 0.05$ & $7.31^{+0.14}_{-0.14}$ & [Ne v] & This work & $0.11^{+0.04}_{-0.04}$ \\
ESO 103- G 035 & 279.58 & -65.43 & 0.013 & / & Sy 2 & ... & $40.82 \pm 0.05$ & $44.93 \pm 0.05$ & $7.71^{+0.10}_{-0.10}$ & [Ne v] & This work & $0.13^{+0.04}_{-0.04}$ \\
ESO 138- G 001 & 252.83 & -59.23 & 0.007 & / & Sy 2 & ... & $40.37 \pm 0.05$ & $44.49 \pm 0.05$ & $7.35^{+0.14}_{-0.14}$ & [Ne v] & This work & $0.11^{+0.04}_{-0.04}$ \\
ESO 140- G 043 & 281.23 & -62.36 & 0.014 & / & Sy 1 & / & $40.53 \pm 0.03$ & $44.64 \pm 0.03$ & $7.47^{+0.12}_{-0.12}$ & [Ne v] & This work & $0.12^{+0.03}_{-0.03}$ \\
ESO 323- G 077 & 196.61 & -40.41 & 0.015 & / & Sy 1.2 & ... & $40.71 \pm 0.03$ & $44.83 \pm 0.03$ & $7.62^{+0.11}_{-0.11}$ & [Ne v] & This work & $0.13^{+0.03}_{-0.03}$ \\
ESO 339- G 011 & 299.41 & -37.94 & 0.019 & $\times$ & Sy 2 & ... & $41.34 \pm 0.03$ & $45.45 \pm 0.03$ & $8.14^{+0.11}_{-0.11}$ & [Ne v] & This work & $0.17^{+0.04}_{-0.04}$ \\
ESO 362- G 008 & 77.79 & -34.39 & 0.016 & $\times$ & Sy 2 & $\times$ & $40.47 \pm 0.03$ & $44.59 \pm 0.03$ & $7.43^{+0.13}_{-0.13}$ & [Ne v] & This work & $0.11^{+0.03}_{-0.03}$ \\
ESO 420- G 013 & 63.46 & -32.01 & 0.012 & $\times$ & Sy 2 & ... & $40.90 \pm 0.04$ & $45.01 \pm 0.04$ & $7.77^{+0.10}_{-0.10}$ & [Ne v] & This work & $0.14^{+0.03}_{-0.03}$ \\
ESO 426- G 002 & 95.94 & -32.22 & 0.022 & / & Sy 2 & / & $40.42 \pm 0.07$ & $44.54 \pm 0.07$ & $7.39^{+0.13}_{-0.13}$ & [Ne v] & This work & $0.11^{+0.04}_{-0.04}$ \\
ESO 428- G 014 & 109.13 & -29.32 & 0.006 & $\times$ & Sy 2 & $\times$ & $40.71 \pm 0.01$ & $44.82 \pm 0.01$ & $7.62^{+0.11}_{-0.11}$ & [Ne v] & This work & $0.13^{+0.03}_{-0.03}$ \\
ESO 434- G 040 & 146.92 & -30.95 & 0.008 & / & Sy 2 & / & $40.42 \pm 0.04$ & $44.53 \pm 0.04$ & $7.39^{+0.13}_{-0.13}$ & [Ne v] & This work & $0.11^{+0.04}_{-0.04}$ \\
ESO 439- G 009 & 171.85 & -29.26 & 0.024 & / & Sy 2 & $\times$ & $41.01 \pm 0.03$ & $45.12 \pm 0.03$ & $7.87^{+0.10}_{-0.10}$ & [Ne v] & This work & $0.14^{+0.03}_{-0.03}$ \\
FAIRALL 0049 & 279.24 & -59.4 & 0.02 & / & NLSy 1 & / & $41.42 \pm 0.02$ & $45.53 \pm 0.02$ & $8.20^{+0.12}_{-0.12}$ & [Ne v] & This work & $0.17^{+0.05}_{-0.05}$ \\
IC 1816 & 37.96 & -36.67 & 0.017 & / & Sy 1 & $\times$ & $40.69 \pm 0.03$ & $44.81 \pm 0.03$ & $7.61^{+0.11}_{-0.11}$ & [Ne v] & This work & $0.13^{+0.03}_{-0.03}$ \\
IC 2560 & 154.08 & -33.56 & 0.01 & $\times$ & Sy 2 & $\times$ & $40.68 \pm 0.04$ & $44.80 \pm 0.04$ & $7.60^{+0.11}_{-0.11}$ & [Ne v] & This work & $0.13^{+0.03}_{-0.03}$ \\
IC 4329A & 207.33 & -30.31 & 0.016 & / & Sy 2 & / & $41.32 \pm 0.02$ & $45.43 \pm 0.02$ & $7.64^{+0.53}_{-0.25}$ & \shortstack{Stellar\\kinematics} & (3) & $0.49^{+0.60}_{-0.28}$ \\
IC 4518A & 224.42 & -43.13 & 0.016 & / & Sy 2 & ... & $41.23 \pm 0.02$ & $45.34 \pm 0.02$ & $8.05^{+0.11}_{-0.11}$ & [Ne v] & This work & $0.16^{+0.04}_{-0.04}$ \\
IC 5063 & 313.01 & -57.07 & 0.011 & / & Sy 1 & / & $40.87 \pm 0.05$ & $44.99 \pm 0.05$ & $7.74^{+0.00}_{-0.00}$ & M-$\sigma^*$ & (4) & $0.14^{+0.02}_{-0.02}$ \\
IRAS 11215-2806 & 171.01 & -28.39 & 0.014 & $\times$ & Sy 2 & / & $40.27 \pm 0.02$ & $44.38 \pm 0.02$ & $7.26^{+0.15}_{-0.15}$ & [Ne v] & This work & $0.10^{+0.04}_{-0.04}$ \\
MCG -01-24-012 & 140.19 & -8.06 & 0.02 & / & Sy 2 & / & $40.38 \pm 0.04$ & $44.50 \pm 0.04$ & $7.16^{+0.12}_{-0.12}$ & M-$\sigma^*$ & (5) & $0.17^{+0.05}_{-0.05}$ \\
MCG -03-34-064 & 200.6 & -16.73 & 0.017 & / & Sy 2 & / & $41.36 \pm 0.06$ & $45.47 \pm 0.06$ & $7.70^{+0.00}_{-0.00}$ & M-$\sigma^*$ & (6) & $0.47^{+0.07}_{-0.07}$ \\
MESSIER 084 & 186.27 & 12.89 & 0.003 & $\times$ & Sy 2 & $\times$ & $38.16 \pm 0.08$ & $42.27 \pm 0.08$ & $5.54^{+0.44}_{-0.44}$ & \shortstack{Gas\\kinematics} & (7) & $0.04^{+0.04}_{-0.04}$ \\
MRK 0915 & 339.19 & -12.55 & 0.024 & / & Sy 1 & $\times$ & $40.99 \pm 0.05$ & $45.11 \pm 0.05$ & $7.85^{+0.10}_{-0.10}$ & [Ne v] & This work & $0.14^{+0.04}_{-0.04}$ \\
NGC 0235A & 10.72 & -23.54 & 0.022 & / & Sy 1.9 & $\times$ & $40.75 \pm 0.05$ & $44.87 \pm 0.05$ & $7.66^{+0.11}_{-0.11}$ & [Ne v] & This work & $0.13^{+0.03}_{-0.03}$ \\
NGC 0262 & 12.2 & 31.96 & 0.015 & / & Sy 2 & / & $40.65 \pm 0.06$ & $44.76 \pm 0.06$ & $7.06^{+0.00}_{-0.00}$ & M-$\sigma^*$ & (8) & $0.40^{+0.05}_{-0.05}$ \\
NGC 0291 & 13.37 & -8.77 & 0.019 & $\times$ & Sy 2 & $\times$ & $40.43 \pm 0.06$ & $44.55 \pm 0.06$ & $7.10^{+0.00}_{-0.00}$ & M-$\sigma^*$ & (1) & $0.22^{+0.03}_{-0.03}$ \\
NGC 0449 & 19.03 & 33.09 & 0.016 & $\times$ & Sy 2 & ... & $40.90 \pm 0.02$ & $45.01 \pm 0.02$ & $7.78^{+0.10}_{-0.10}$ & [Ne v] & This work & $0.14^{+0.03}_{-0.03}$ \\
NGC 0526A & 20.98 & -35.07 & 0.019 & / & Sy 1 & / & $40.58 \pm 0.06$ & $44.69 \pm 0.06$ & $7.51^{+0.12}_{-0.12}$ & [Ne v] & This work & $0.12^{+0.04}_{-0.04}$ \\
NGC 0788 & 30.28 & -6.82 & 0.014 & / & Sy 1.8 & / & $40.46 \pm 0.02$ & $44.57 \pm 0.02$ & $7.06^{+0.00}_{-0.00}$ & M-$\sigma^*$ & (8) & $0.26^{+0.01}_{-0.01}$ \\
NGC 0973 & 38.58 & 32.51 & 0.016 & / & Sy 1 & $\times$ & $40.28 \pm 0.06$ & $44.39 \pm 0.06$ & $7.27^{+0.15}_{-0.15}$ & [Ne v] & This work & $0.10^{+0.04}_{-0.04}$ \\
NGC 1365 & 53.4 & -36.14 & 0.005 & / & Sy 1 & $\times$ & $40.47 \pm 0.07$ & $44.59 \pm 0.07$ & $7.29^{+0.79}_{-0.79}$ & FWHM (Ha) & (9) & $0.16^{+0.28}_{-0.28}$ \\
NGC 1386 & 54.19 & -36.0 & 0.003 & $\times$ & Sy 2 & $\times$ & $39.91 \pm 0.01$ & $44.02 \pm 0.01$ & $7.24^{+0.00}_{-0.00}$ & M-$\sigma^*$ & (4) & $0.05^{+0.00}_{-0.00}$ \\
NGC 2110 & 88.05 & -7.46 & 0.008 & / & Sy 2 & / & $39.98 \pm 0.08$ & $44.09 \pm 0.08$ & $8.18^{+13.50}_{-13.50}$ & M-$\sigma^*$ & (9) & $0.01^{+0.20}_{-0.20}$ \\
NGC 2273 & 102.54 & 60.85 & 0.006 & / & Sy 1.9 & $\times$ & $39.79 \pm 0.05$ & $43.91 \pm 0.05$ & $6.88^{+0.21}_{-0.21}$ & [Ne v] & This work & $0.09^{+0.04}_{-0.04}$ \\
NGC 2623 & 129.6 & 25.75 & 0.019 & $\times$ & Composite & $\times$ & $40.52 \pm 0.08$ & $44.63 \pm 0.08$ & $7.47^{+0.12}_{-0.12}$ & [Ne v] & This work & $0.12^{+0.04}_{-0.04}$ \\
NGC 2992 & 146.43 & -14.33 & 0.008 & / & Sy 2 & $\times$ & $40.59 \pm 0.03$ & $44.71 \pm 0.03$ & $7.36^{+0.92}_{-0.92}$ & FWHM (Ha) & (9) & $0.18^{+0.37}_{-0.37}$ \\
NGC 3081 & 149.87 & -22.83 & 0.008 & / & Sy 2 & $\times$ & $40.69 \pm 0.02$ & $44.81 \pm 0.02$ & $7.56^{+3.56}_{-3.56}$ & M-$\sigma^*$ & (9) & $0.14^{+1.14}_{-1.14}$ \\
NGC 3227 & 155.88 & 19.87 & 0.004 & / & Sy 2 & $\times$ & $39.94 \pm 0.04$ & $44.05 \pm 0.04$ & $7.06^{+0.00}_{-0.00}$ & RM & (2) & $0.08^{+0.01}_{-0.01}$ \\
NGC 3281 & 157.97 & -34.85 & 0.011 & / & Sy 2 & / & $40.93 \pm 0.16$ & $45.04 \pm 0.16$ & $7.80^{+0.10}_{-0.10}$ & [Ne v] & This work & $0.14^{+0.06}_{-0.06}$ \\
NGC 3374 & 162.0 & 43.19 & 0.025 & $\times$ & Not AGN & $\times$ & $40.25 \pm 0.11$ & $44.36 \pm 0.11$ & $7.24^{+0.15}_{-0.15}$ & [Ne v] & This work & $0.10^{+0.04}_{-0.04}$ \\
NGC 3393 & 162.1 & -25.16 & 0.013 & / & Sy 1 & $\times$ & $41.12 \pm 0.01$ & $45.24 \pm 0.01$ & $7.96^{+0.10}_{-0.10}$ & [Ne v] & This work & $0.15^{+0.04}_{-0.04}$ \\
NGC 3735 & 173.99 & 70.54 & 0.009 & $\times$ & Sy 2 & $\times$ & $40.18 \pm 0.04$ & $44.30 \pm 0.04$ & $7.19^{+0.16}_{-0.16}$ & [Ne v] & This work & $0.10^{+0.04}_{-0.04}$ \\
NGC 3783 & 174.76 & -37.74 & 0.01 & / & Sy 1 & / & $40.55 \pm 0.04$ & $44.66 \pm 0.04$ & $7.15^{+0.11}_{-0.12}$ & RM & (1) & $0.26^{+0.07}_{-0.08}$ \\
NGC 3786 & 174.93 & 31.91 & 0.009 & / & Sy 1.8 & $\times$ & $39.80 \pm 0.04$ & $43.91 \pm 0.04$ & $6.88^{+0.21}_{-0.21}$ & [Ne v] & This work & $0.09^{+0.04}_{-0.04}$ \\
NGC 4074 & 181.12 & 20.32 & 0.022 & / & Sy 1.2 & $\times$ & $40.89 \pm 0.03$ & $45.01 \pm 0.03$ & $7.58^{+0.00}_{-0.00}$ & M-$\sigma^*$ & (8) & $0.21^{+0.01}_{-0.01}$ \\
NGC 4151 & 182.64 & 39.41 & 0.003 & / & Sy 1.8 & / & $40.40 \pm 0.02$ & $44.51 \pm 0.02$ & $7.58^{+0.00}_{-0.00}$ & RM & (2) & $0.07^{+0.00}_{-0.00}$ \\
NGC 4388 & 186.44 & 12.66 & 0.008 & / & Sy 2 & / & $40.71 \pm 0.06$ & $44.82 \pm 0.06$ & $8.40^{+0.20}_{-0.20}$ & Megamaser & (2) & $0.02^{+0.01}_{-0.01}$ \\
NGC 4395 & 186.45 & 33.55 & 0.001 & / & Sy 1.8 & $\times$ & $37.47 \pm 0.03$ & $41.58 \pm 0.03$ & $4.98^{+0.54}_{-0.54}$ & [Ne v] & This work & $0.03^{+0.04}_{-0.04}$ \\
NGC 4507 & 188.9 & -39.91 & 0.012 & / & Sy 1 & / & $40.48 \pm 0.07$ & $44.59 \pm 0.07$ & $7.43^{+0.13}_{-0.13}$ & [Ne v] & This work & $0.11^{+0.04}_{-0.04}$ \\
NGC 4818 & 194.2 & -8.53 & 0.004 & $\times$ & Starburst & $\times$ & $39.24 \pm 0.11$ & $43.36 \pm 0.11$ & $6.43^{+0.28}_{-0.28}$ & [Ne v] & This work & $0.07^{+0.05}_{-0.05}$ \\
NGC 4939 & 196.06 & -10.34 & 0.01 & / & Sy 2 & $\times$ & $40.51 \pm 0.02$ & $44.62 \pm 0.02$ & $7.46^{+0.12}_{-0.12}$ & [Ne v] & This work & $0.12^{+0.03}_{-0.03}$ \\
NGC 4941 & 196.05 & -5.55 & 0.004 & / & Sy 2 & $\times$ & $39.37 \pm 0.03$ & $43.48 \pm 0.03$ & $6.53^{+0.26}_{-0.26}$ & [Ne v] & This work & $0.07^{+0.04}_{-0.04}$ \\
NGC 5005 & 197.73 & 37.06 & 0.003 & $\times$ & LINER & ... & $38.91 \pm 0.14$ & $43.03 \pm 0.14$ & $6.16^{+0.33}_{-0.33}$ & [Ne v] & This work & $0.06^{+0.05}_{-0.05}$ \\
NGC 5128 & 201.37 & -43.02 & 0.002 & / & Sy 2 & ... & $39.35 \pm 0.05$ & $43.46 \pm 0.05$ & $6.51^{+0.27}_{-0.27}$ & \shortstack{Gas\\kinematics} & (11) & $0.07^{+0.04}_{-0.04}$ \\
NGC 5135 & 201.43 & -29.83 & 0.014 & $\times$ & Sy 2 & $\times$ & $41.00 \pm 0.04$ & $45.12 \pm 0.04$ & $7.35^{+0.00}_{-0.00}$ & M-$\sigma^*$ & (4) & $0.46^{+0.04}_{-0.04}$ \\
NGC 5227 & 203.85 & 1.41 & 0.017 & $\times$ & Sy 2 & $\times$ & $39.78 \pm 0.07$ & $43.89 \pm 0.07$ & $6.86^{+0.21}_{-0.21}$ & [Ne v] & This work & $0.08^{+0.04}_{-0.04}$ \\
NGC 5252 & 204.57 & 4.54 & 0.023 & / & Sy 1.9 & $\times$ & $40.20 \pm 0.05$ & $44.31 \pm 0.05$ & $7.21^{+0.16}_{-0.16}$ & \shortstack{Gas\\kinematics} & (12) & $0.10^{+0.04}_{-0.04}$ \\
NGC 5347 & 208.32 & 33.49 & 0.008 & $\times$ & Sy 2 & / & $39.78 \pm 0.08$ & $43.90 \pm 0.08$ & $6.79^{+0.00}_{-0.00}$ & M-$\sigma^*$ & (4) & $0.10^{+0.02}_{-0.02}$ \\
NGC 5427 & 210.86 & -6.03 & 0.009 & $\times$ & Sy 2 & $\times$ & $39.51 \pm 0.38$ & $43.62 \pm 0.38$ & $6.65^{+0.24}_{-0.24}$ & [Ne v] & This work & $0.08^{+0.08}_{-0.08}$ \\
NGC 5506 & 213.31 & -3.21 & 0.006 & / & Sy 2 & / & $40.76 \pm 0.03$ & $44.88 \pm 0.03$ & $7.35^{+2.34}_{-2.34}$ & M-$\sigma^*$ & (9) & $0.27^{+1.45}_{-1.45}$ \\
NGC 5548 & 214.5 & 25.14 & 0.017 & / & Sy 1.5 & ... & $40.46 \pm 0.07$ & $44.58 \pm 0.07$ & $8.09^{+0.12}_{-0.12}$ & RM & (10) & $0.02^{+0.01}_{-0.01}$ \\
NGC 5643 & 218.17 & -44.17 & 0.004 & / & Sy 2 & $\times$ & $39.98 \pm 0.03$ & $44.10 \pm 0.03$ & $6.44^{+0.11}_{-0.21}$ & M-$\sigma^*$ & (13) & $0.36^{+0.09}_{-0.18}$ \\
NGC 5695 & 219.34 & 36.57 & 0.014 & $\times$ & Sy 2 & $\times$ & $39.99 \pm 0.06$ & $44.10 \pm 0.06$ & $7.44^{+0.00}_{-0.00}$ & M-$\sigma^*$ & (1) & $0.04^{+0.01}_{-0.01}$ \\
NGC 5728 & 220.6 & -17.25 & 0.009 & / & Sy 2 & $\times$ & $40.50 \pm 0.05$ & $44.61 \pm 0.05$ & $7.36^{+0.31}_{-0.31}$ & Megamaser & (9) & $0.14^{+0.10}_{-0.10}$ \\
NGC 5899 & 228.76 & 42.05 & 0.009 & / & Sy 2 & $\times$ & $40.00 \pm 0.04$ & $44.11 \pm 0.04$ & $7.04^{+0.18}_{-0.18}$ & [Ne v] & This work & $0.09^{+0.04}_{-0.04}$ \\
NGC 6221 & 253.19 & -59.22 & 0.005 & / & Sy 2 & $\times$ & $39.87 \pm 0.13$ & $43.98 \pm 0.13$ & $6.46^{+0.06}_{-0.06}$ & M-$\sigma^*$ & (5) & $0.27^{+0.09}_{-0.09}$ \\
NGC 6240 & 253.25 & 2.4 & 0.024 & / & Sy 2 & $\times$ & $40.99 \pm 0.14$ & $45.10 \pm 0.14$ & $7.85^{+0.10}_{-0.10}$ & [Ne v] & This work & $0.14^{+0.06}_{-0.06}$ \\
NGC 6300 & 259.25 & -62.82 & 0.004 & / & Sy 2 & / & $39.64 \pm 0.03$ & $43.75 \pm 0.03$ & $6.80^{+0.11}_{-0.22}$ & M-$\sigma^*$ & (14) & $0.07^{+0.02}_{-0.04}$ \\
NGC 6328 & 260.92 & -65.01 & 0.014 & $\times$ & LINER & ... & $39.32 \pm 0.12$ & $43.44 \pm 0.12$ & $6.49^{+0.27}_{-0.27}$ & [Ne v] & This work & $0.07^{+0.05}_{-0.05}$ \\
NGC 6764 & 287.07 & 50.93 & 0.008 & $\times$ & LINER & $\times$ & $39.72 \pm 0.07$ & $43.83 \pm 0.07$ & $6.82^{+0.22}_{-0.22}$ & [Ne v] & This work & $0.08^{+0.04}_{-0.04}$ \\
NGC 6814 & 295.67 & -10.32 & 0.005 & / & Sy 1.5 & $\times$ & $39.57 \pm 0.06$ & $43.69 \pm 0.06$ & $7.07^{+0.11}_{-0.11}$ & RM & (10) & $0.03^{+0.01}_{-0.01}$ \\
NGC 6926 & 308.28 & -2.03 & 0.02 & $\times$ & Sy 2 & $\times$ & $40.23 \pm 0.13$ & $44.35 \pm 0.13$ & $7.23^{+0.15}_{-0.15}$ & [Ne v] & This work & $0.10^{+0.05}_{-0.05}$ \\
NGC 6951 & 309.31 & 66.11 & 0.005 & $\times$ & Sy 2 & $\times$ & $39.35 \pm 0.13$ & $43.47 \pm 0.13$ & $6.52^{+0.27}_{-0.27}$ & [Ne v] & This work & $0.07^{+0.05}_{-0.05}$ \\
NGC 7130 & 327.08 & -34.95 & 0.016 & / & Composite & $\times$ & $40.94 \pm 0.05$ & $45.05 \pm 0.05$ & $7.53^{+0.00}_{-0.00}$ & M-$\sigma^*$ & (4) & $0.26^{+0.03}_{-0.03}$ \\
NGC 7172 & 330.51 & -31.87 & 0.009 & / & Sy 2 & / & $40.36 \pm 0.03$ & $44.47 \pm 0.03$ & $7.73^{+0.51}_{-0.51}$ & M-$\sigma^*$ & (9) & $0.04^{+0.05}_{-0.05}$ \\
NGC 7314 & 338.94 & -26.05 & 0.005 & / & Sy 1.9 & $\times$ & $39.93 \pm 0.02$ & $44.04 \pm 0.02$ & $6.24^{+0.00}_{-0.00}$ & RM & (5) & $0.50^{+0.03}_{-0.03}$ \\
NGC 7469 & 345.82 & 8.87 & 0.016 & / & Sy 1.2 & $\times$ & $41.16 \pm 0.04$ & $45.28 \pm 0.04$ & $7.69^{+0.13}_{-0.13}$ & RM & (10) & $0.31^{+0.10}_{-0.10}$ \\
NGC 7682 & 352.27 & 3.53 & 0.017 & / & Sy 2 & $\times$ & $40.39 \pm 0.03$ & $44.50 \pm 0.03$ & $7.36^{+0.14}_{-0.14}$ & [Ne v] & This work & $0.11^{+0.04}_{-0.04}$ \\
PGC 038055 & 180.74 & -20.93 & 0.022 & $\times$ & Sy 1.8 & $\times$ & $39.90 \pm 0.03$ & $44.02 \pm 0.03$ & $6.96^{+0.19}_{-0.19}$ & [Ne v] & This work & $0.09^{+0.04}_{-0.04}$ \\
Phoenix Galaxy & 121.02 & 5.11 & 0.014 & / & Sy 2 & / & $40.83 \pm 0.02$ & $44.94 \pm 0.02$ & $7.72^{+0.10}_{-0.10}$ & [Ne v] & This work & $0.13^{+0.03}_{-0.03}$ \\
UGC 00006 & 0.79 & 21.96 & 0.022 & $\times$ & Sy 1.8 & $\times$ & $40.87 \pm 0.03$ & $44.99 \pm 0.03$ & $7.75^{+0.10}_{-0.10}$ & [Ne v] & This work & $0.14^{+0.03}_{-0.03}$ \\
UGC 00987 & 21.38 & 32.14 & 0.016 & $\times$ & Sy 2 & $\times$ & $40.12 \pm 0.09$ & $44.23 \pm 0.09$ & $7.14^{+0.17}_{-0.17}$ & [Ne v] & This work & $0.10^{+0.04}_{-0.04}$ \\
UGC 01214 & 25.99 & 2.35 & 0.017 & $\times$ & Sy 2 & / & $41.25 \pm 0.02$ & $45.36 \pm 0.02$ & $7.28^{+0.00}_{-0.00}$ & M-$\sigma^*$ & (8) & $0.95^{+0.03}_{-0.03}$ \\
UGC 01395 & 28.84 & 6.61 & 0.017 & $\times$ & Sy 1.9 & $\times$ & $39.99 \pm 0.07$ & $44.11 \pm 0.07$ & $7.04^{+0.18}_{-0.18}$ & [Ne v] & This work & $0.09^{+0.04}_{-0.04}$ \\
UGC 02456 & 44.99 & 36.82 & 0.012 & / & Sy 2 & $\times$ & $40.48 \pm 0.08$ & $44.59 \pm 0.08$ & $7.01^{+0.00}_{-0.00}$ & M-$\sigma^*$ & (8) & $0.30^{+0.05}_{-0.05}$ \\
UGC 02608 & 48.76 & 42.04 & 0.023 & $\times$ & Sy 2 & / & $41.58 \pm 0.03$ & $45.69 \pm 0.03$ & $8.33^{+0.13}_{-0.13}$ & [Ne v] & This work & $0.18^{+0.06}_{-0.06}$ \\
UGC 03094 & 68.89 & 19.17 & 0.025 & $\times$ & Unclassified & $\times$ & $40.71 \pm 0.08$ & $44.82 \pm 0.08$ & $7.62^{+0.11}_{-0.11}$ & [Ne v] & This work & $0.13^{+0.04}_{-0.04}$ \\
UGC 03426 & 93.9 & 71.04 & 0.014 & / & Sy 2 & ... & $41.45 \pm 0.02$ & $45.56 \pm 0.02$ & $8.65^{+0.00}_{-0.00}$ & M-$\sigma^*$ & (8) & $0.06^{+0.00}_{-0.00}$ \\
UGC 03478 & 98.2 & 63.67 & 0.013 & / & Sy 1.2 & / & $40.60 \pm 0.03$ & $44.72 \pm 0.03$ & $7.53^{+0.12}_{-0.12}$ & [Ne v] & This work & $0.12^{+0.03}_{-0.03}$ \\
UGC 03601 & 103.96 & 40.0 & 0.017 & / & Sy 1.5 & $\times$ & $40.22 \pm 0.05$ & $44.33 \pm 0.05$ & $7.22^{+0.15}_{-0.15}$ & [Ne v] & This work & $0.10^{+0.04}_{-0.04}$ \\
UGC 04229 & 121.92 & 39.0 & 0.023 & / & Sy 2 & $\times$ & $40.47 \pm 0.06$ & $44.58 \pm 0.06$ & $6.92^{+0.00}_{-0.00}$ & M-$\sigma^*$ & (8) & $0.36^{+0.05}_{-0.05}$ \\
UGC 07064 & 181.18 & 31.18 & 0.025 & / & Sy 1.9 & $\times$ & $40.86 \pm 0.04$ & $44.98 \pm 0.04$ & $7.75^{+0.10}_{-0.10}$ & [Ne v] & This work & $0.13^{+0.03}_{-0.03}$ \\
UGC 09883 & 232.84 & 47.02 & 0.021 & $\times$ & Starburst & ... & $40.03 \pm 0.15$ & $44.15 \pm 0.15$ & $7.07^{+0.18}_{-0.18}$ & [Ne v] & This work & $0.09^{+0.05}_{-0.05}$ \\
UGC 12138 & 340.07 & 8.05 & 0.025 & / & Sy 1.8 & $\times$ & $40.86 \pm 0.04$ & $44.97 \pm 0.04$ & $7.74^{+0.10}_{-0.10}$ & [Ne v] & This work & $0.13^{+0.03}_{-0.03}$ \\
UGC 12914 & 0.41 & 23.48 & 0.015 & $\times$ & Composite & $\times$ & $39.72 \pm 0.10$ & $43.83 \pm 0.10$ & $6.81^{+0.22}_{-0.22}$ & [Ne v] & This work & $0.08^{+0.05}_{-0.05}$ \\
UGC8387 & 200.15 & 34.14 & 0.023 & $\times$ & Unclassified & $\times$ & $40.95 \pm 0.07$ & $45.06 \pm 0.07$ & $7.82^{+0.10}_{-0.10}$ & [Ne v] & This work & $0.14^{+0.04}_{-0.04}$ \\
\enddata


\tablecomments{
Columns (2)–(4) Source position and redshift.
Columns (5)–(7) AGN detection in the X-ray (\textit{Swift}/BAT), optical (nuclear activity type), and infrared (\textit{WISE}) bands, respectively.
Columns (8)–(9) list the $\rm [Ne\,v]$ and bolometric luminosities, with $L_{\rm bol}$ derived from $L_{\rm [Ne\,v]}$.
Columns (10)–(12) Black hole mass, method used to derive it, and the reference.
Column (13) Eddington ratio.
For columns (5)–(7), “/” denotes AGN, “$\times$” denotes non-AGN, and “...” indicates unavailable data. References for black hole mass determination are provided below.}

\tablerefs{(1) \citealt{lamassa10};
(2) \citealt{bentz23};
(3) \citealt{winkel24};
(4) \citealt{biangu07};
(5) \citealt{onori17};
(6) \citealt{peng06};
(7) \citealt{walsh10};
(8) \citealt{woo02};
(9) \citealt{caglar20};
(10) \citealt{ho_kim15};
(11) \citealt{gnerucci11};
(12) \citealt{capetti05};
(13) \citealt{whittle92};
(14) \citealt{goulding10}}

\end{deluxetable}
\end{longrotatetable}

\bibliography{references}{}
\bibliographystyle{aasjournalv7}



\end{document}